\documentclass[aps,prl,twocolumn,showpacs,superscriptaddress,amsmath,amssymb]{revtex4-2}
\usepackage{graphicx}
\usepackage{latexsym}
\usepackage{amsmath,amssymb}
\usepackage{bm} 
\usepackage{color}
\usepackage{epsfig}
\usepackage{multirow}
\usepackage{xcolor}
\usepackage{colortbl}
\usepackage{hhline}

\usepackage{graphicx}
\usepackage[colorlinks=true,linkcolor=blue,citecolor=blue,urlcolor=blue]{hyperref}
\usepackage{bbold}
\usepackage{gensymb}

\usepackage{pgfplots,pgfplotstable}
\tikzset{every axis plot post/.append style={solid, thin},every mark/.append style={solid,scale=1}}
\usetikzlibrary{decorations.markings}
\tikzset{mdar/.style ={decoration={markings,mark={at position 0.5 with {\fill (2pt,0)--(-2pt,2pt)--(-2pt,-2pt)--cycle;}}},postaction={decorate}}}
\tikzstyle{dsh}=[dash pattern=on 2.5pt off 1.5pt]
\tikzset{intcur/.style ={decorate,decoration={snake,amplitude=.5mm,segment length=3mm}}}
\tikzstyle{zba}=[dash pattern=on 1pt off 1pt,line width=2pt]

\usepackage{simplewick}

\definecolor{myblue}{rgb}{.93, .93, 1}

\AtBeginDocument{%
    \newwrite\bibnotes
    \def\bibnotesext{Notes.bib}
    \immediate\openout\bibnotes=\jobname\bibnotesext
    \immediate\write\bibnotes{@CONTROL{REVTEX42Control}}
    \immediate\write\bibnotes{@CONTROL{%
    apsrev42Control,author="08",editor="1",pages="0",title="0",year="1"}}
     \if@filesw
     \immediate\write\@auxout{\string\citation{apsrev42Control}}%
    \fi
}%

\def \a {\alpha}
\def \b {\beta}
\def \d {\delta}
\def \D {\Delta}

\def \ve {\varepsilon}

\def \G{\Gamma}

\def \l {\lambda}
\def \L {\Lambda}

\def \s {\sigma}

\def \dag {\dagger}

\def \p {\partial}

\def \eqv {\equiv}
\def \apx {\approx}

\def \til {\tilde}

\def \dag {\dagger}

\newcommand{\intv}[1]{\int_{\mbf #1}}

\newcommand{\sumv}[1]{\sum_{\mbf #1}}

\def \rar {\rightarrow}

\def \la {\langle}
\def \ra {\rangle}
\def \fr {\frac}
\def \lf {\left}
\def \ri {\right}

\newcommand{\epvl}[1]{\la#1\ra}

\def \bece {\begin{center}}
\def \ence {\end{center}}
\def \beeq {\begin{equation}}
\def \eneq {\end{equation}}
\def \beal {\begin{aligned}}
\def \enal {\end{aligned}}
\def \bega {\begin{gathered}}
\def \enga {\end{gathered}}
\def \benu {\begin{enumerate}}
\def \ennu {\end{enumerate}}
\def \beit {\begin{itemize}}
\def \enit {\end{itemize}}
\def \bede {\begin{description}}
\def \ende {\end{description}}
\def \betb {\begin{tabular}}
\def \entb {\end{tabular}}
\def \bear {\begin{array}}
\def \enar {\end{array}}

\def \mbf {\mathbf}
\def \mbb {\mathbb}

\def \bsb{\boldsymbol}
\def \txt {\text}

\newcommand{\comment}[1]{}

\begin{document}


\title{Multidome superconductivity in charge density wave kagome metals}

\author{Yu-Ping Lin}
\affiliation{Department of Physics, University of Colorado, Boulder, Colorado 80309, USA}
\author{Rahul M. Nandkishore}
\affiliation{Department of Physics, University of Colorado, Boulder, Colorado 80309, USA}
\affiliation{Center for Theory of Quantum Matter, University of Colorado, Boulder, Colorado 80309, USA}

\date{\today}

\begin{abstract}
Motivated by recent experiments on the kagome metals $A\text{V}_3\text{Sb}_5$ with $A=\text{K}$, $\text{Rb}$, and $\text{Cs}$, which show a charge density wave (CDW) at $\sim100$ K and the superconductivity at $\sim1$ K, we explore the onset of the superconductivity, taking the perspective that it descends from a parent CDW. We argue that viewing the superconductivity as a weak-coupling instability of a reconstructed (by the CDW) Fermi surface naturally explains the experimentally observed `multidome' nonmonotonic dependence on pressure, with the `peaks' in the superconducting critical temperature being associated with the Van Hove singularities of the reconstructed Fermi surface. This `parent-child relationship' also naturally explains the large separation of energy scales between the superconductivity and the CDW. We discuss different possible pairing mechanisms and speculate that the CDW or reconstructed Pomeranchuk fluctuations may mediate the pairing interaction.
\end{abstract}

\maketitle


\textit{Introduction.---}The correlated phenomena in the recently uncovered kagome metals $A\txt{V}_3\txt{Sb}_5$ with $A=\txt{K}$, $\txt{Rb}$, and $\txt{Cs}$ have drawn enormous attention. Charge density waves (CDWs) were observed at high critical temperature $T_\txt{CDW}\sim100\txt{ K}$ \cite{ortiz19prm,yang20sa,ortiz20prl,kenney21jpcm,jiang21nm,chen21prl,yu21prb,du21prb,zhao21n,liang21prx,uykur22npjqm,chen21n,li21prx,wang21ax,nakayama21prb,li22np,shumiya21prb,tsirlin21sp,kang22np,wang21prb,cho21prl,song21ax,song21prl,yu21nc,hu22nc,mielke22n,wang21prr,luo22nc,qian21prb}. With simultaneous ordering at three commensurate momenta, these $3Q$ orders are believed to be driven by the Fermi surface nesting, further enhanced by the Van Hove singularity (VHS) \cite{vanhove53pr}. The bond density modulations form (inverse) star-of-David patterns \cite{tan21prl,feng21sb,denner21prl,lin21prb,park21prb,setty21ax,feng21prb,miao21prb,christensen21prb} with possible higher-order topology \cite{lin21ax}. Meanwhile, the giant anomalous Hall effects indicate a band topology with time-reversal symmetry breaking \cite{yang20sa,yu21prb}, which may be attributed to the loop currents \cite{affleck88prb,varma,nayak00prb,chakravarty01prb,venderbos16prb,lin19prb, jiang21nm,shumiya21prb,wang21prb,mielke22n,feng21sb,denner21prl,lin21prb,park21prb,feng21prb}. On the other hand, the $\txt{C}_3$ symmetry breaking is generally observed and is possibly induced by the three-dimensional orders \cite{park21prb,miao21prb,christensen21prb}.

More exotic observations appeared in the superconductivity (SC) \cite{ortiz20prl,chen21prl,du21prb,liang21prx,chen21n,song21ax,song21prl,yu21nc,wang21prr,qian21prb,ortiz21prm,wang20ax,zhao21axsc,duan21scp,zhang21prb,ni21cpl,xiang21nc,xu21prl,zhu22prb}. The superconductivity develops at much lower critical temperature $T_\txt{SC}\sim1\txt{ K}$ with the $\txt{C}_3$ symmetry breaking persisting \cite{ni21cpl,xiang21nc}. Remarkably, the superconductivity exhibits a double-dome structure under pressure, which extends above the critical pressure of the CDW \cite{chen21prl,yu21nc,du21prb,wang21prr}. New superconducting domes can even appear at much higher pressure \cite{du21prb,zhang21prb,zhu22prb}. Despite the rich experimental observations, a thorough theoretical understanding remains elusive. Given the proximity to the VHS, the superconductivity may be treated as a competing instability with the CDW \cite{wu21prl}. This scenario may explain the opposite trends of $T_\txt{CDW}$ and $T_\txt{SC}$ in certain experimental regimes, such as under pressure \cite{chen21prl,du21prb,zhang21prb,yu21nc,zhu22prb,wang21prr} and uniaxial strain \cite{qian21prb}. However, the scenario does not fit the large separation of energy scales between the superconductivity and the CDW, nor does it explain the double-dome superconductivity under the monotonically suppressed CDW. An alternative scenario is thus suggested for the origin of the superconductivity.

\begin{figure}[b]
\centering
\includegraphics[scale = 0.6]{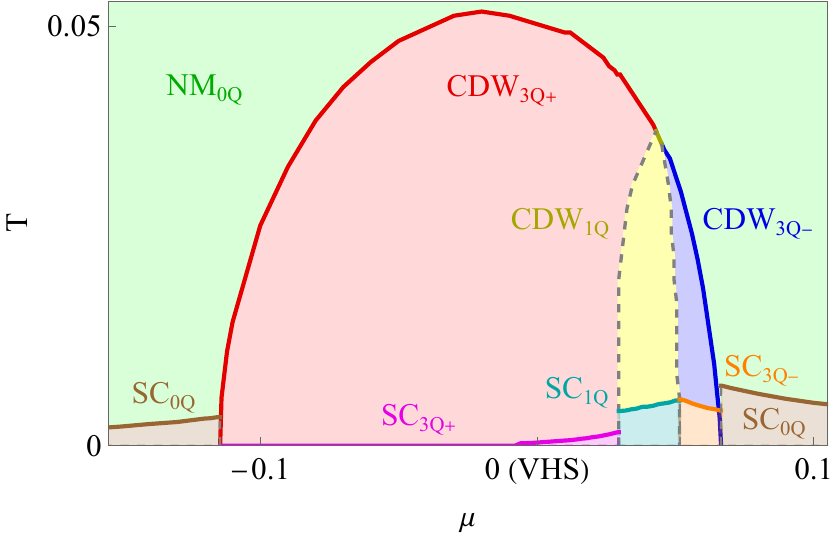}
\caption{\label{fig:pd} Phase diagram of the parent CDW and the multidome child superconductivity. The CDW is computed from the free energy (\ref{eq:fe}) with $\L_k=1$. The dispersion coefficients $A_+=3.15$ and $A_-=1.45$ are set according to the tight-binding model for Figs.~\ref{fig:setup} and \ref{fig:rfs}. The interaction $V_\txt{CDW}=25$ is chosen to approach the large gap in the experiment \cite{kang22np}. The critical temperature of the superconductivity $T_\txt{SC}$ [Eqs.~(\ref{eq:tc}) and (\ref{eq:dlint})] is estimated with $\L_\txt{SC}=0.01$ and $C_\txt{SC}=2$.}
\end{figure}

In this Research Letter, we propose an alternative scenario that appears consistent with the salient experimental facts. We adopt the perspective that the CDW is a `parent' phase, and that the `child' superconductivity emerges from the reconstructed Fermi surface thereof. This `parent-child relationship' not only naturally explains the large separation of energy scales between the superconductivity and the CDW but also predicts a multidome structure for the superconductivity (Fig.~\ref{fig:pd}) \cite{chen21prl,yu21nc,wang21prr}, with the critical temperature enhanced near the (reconstructed) VHSs. We further discuss possible pairing mechanisms and speculate that the CDW or reconstructed Pomeranchuk (RPOM) fluctuations may mediate the pairing interaction.


\begin{figure}[b]
\centering
\includegraphics[scale = 1]{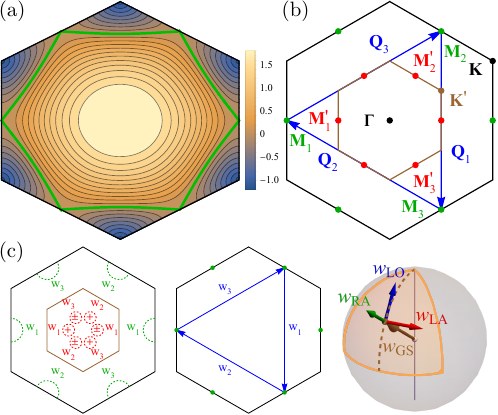}
\caption{\label{fig:setup} Model setup. (a) The middle band of the kagome lattice with the (green) VH Fermi surface in the (black) BZ. In addition to the nearest-neighbor hopping $t_1=1$, we introduce an intrasublattice hopping $t_2=0.05$ between next-nearest-neighbor unit cells to induce the nonperfect nesting. (b) High-symmetry points and nesting momenta in the (black) original and (brown) reduced BZs. (c) In the patch models (left) and at the nesting momenta (middle), the symmetry channels can be mapped onto a Bloch sphere (right).}
\end{figure}

\textit{CDW kagome metal.---}Our analysis focuses on the Van Hove (VH) CDW on the middle band of the kagome lattice. At the VHS, the Fermi surface is a hexagon in the hexagonal Brillouin zone (BZ). The corner saddle points sit at the zone edge centers $\mbf M_{\a=1,2,3}$ [Figs.~\ref{fig:setup}(a) and \ref{fig:setup}(b)], where the particle-type dispersion energy reads $\ve_{\a\mbf k}=A_+k_{\a+}^2-A_-k_{\a-}^2$ with $A_\pm>0$ for small relative momenta $(k_{\a+},k_{\a-})$ from $\mbf M_\a$. The density of states experiences the logarithmic VHS at these points \cite{vanhove53pr}, which enhances the Fermi liquid instabilities in various channels \cite{nandkishore12np,kiesel12prb,yu12prb,kiesel13prl,wang13prb,nandkishore14prb,lin19prb,classen20prb,lin20prb,park21prb}. In particular, the Fermi surface nesting can enhance the CDW at three commensurate nesting momenta $\mbf Q_\a\eqv\mbf M_\a$. Due to the doubled periodicity $2\mbf Q_\a\eqv0$, the bands are folded quadruply onto the $1/2\times1/2$ reduced BZ. The CDW then opens the gaps and shrinks the reconstructed Fermi surface.

We choose the real CDW $\vec\D=(\D_1,\D_2,\D_3)$ specifically to exemplify our analysis. The energetically favored ground states can be determined by minimizing the mean-field free energy \cite{colemanmbp}
\beeq
\label{eq:fe}
f[\vec\D]=\fr{2}{V_\txt{CDW}}-\sum_n\intv{k}\ln[1+e^{-(E_{n\mbf k}[\vec\D]-\mu)/T}].
\eneq
Each energy $E_{n\mbf k}[\vec\D]$ is an eigenvalue of the effective Hamiltonian
\beeq
\label{eq:hcdw3band}
H_{\txt{CDW},\mbf k}[\vec\D]=\lf(\bear{ccc}\ve_{1\mbf k}&-\D_3&-\D_2\\-\D_3&\ve_{2\mbf k}&-\D_1\\-\D_2&-\D_1&\ve_{3\mbf k}\enar\ri),
\eneq
which characterizes a low-energy patch model with the radial cutoff $\L_k$ around each saddle point [Fig.~\ref{fig:setup}(c)]. $V_\txt{CDW}$ defines the projected interaction in the CDW channel. The phase diagram (Fig.~\ref{fig:pd}) \cite{park21prb} is consistent with the intuition from the Ginzburg-Landau theory \cite{denner21prl,lin21prb,park21prb}. Near the VHS, the CDW expands a strong $3Q+$ phase with high critical temperature $T_\txt{CDW}$. These $3Q+$ orders develop the simultaneous ordering $|\D_1|=|\D_2|=|\D_3|$ under $\D_1\D_2\D_3>0$, which maximizes the gap structures by opening the gaps at the Fermi level \cite{venderbos16prb,lin21prb}. In practical systems, the Fermi level usually lies away from the VHS. The nonzero doping $\mu\neq0$ can draw the reconstructed Fermi surface and alter the energetic favor. While the $3Q+$ orders remain stable at negative doping $\mu<0$, the $3Q-$ orders with $\D_1\D_2\D_3<0$ can emerge at positive doping $\mu>0$. Near the phase transition, the CDW is pinned to the $1Q$ orders at single momenta. The nonzero doping suppresses the CDW phase and reduces the critical temperature $T_\txt{CDW}$ gradually. This is attributed to the reduction of the density of states and Fermi surface nesting. When the CDW vanishes at the critical dopings, the ground state returns to the $0Q$ normal metal (NM). Note that different phases are separated by the first-order transitions, where the ground state switches abruptly between different orders. While the phase diagram has been presented previously \cite{park21prb}, we further show that it contains several properties essential to the multidome superconductivity.

Remarkably, the CDW phase can be suppressed by different variations in the kagome metals $A\txt{V}_3\txt{Sb}_5$. In addition to the nonzero doping \cite{song21prl}, the suppression was also observed under pressure \cite{du21prb,chen21prl,yu21nc,wang21prr} and uniaxial strain \cite{qian21prb}. These effects may be related to the Fermi level shifting \cite{chen21prl} or the interaction reduction under structural variation \cite{tsirlin21sp}. In the low-energy theory, the variations can be summarized by the tunings of doping, Fermi surface nesting, and interaction strength. We model these effects by a single parameter, which defines the shifting away from the perfectly nested VHS at a fixed dimensionless interaction. The shifting further determines the reduction of the critical temperature $T_\txt{CDW}$. Our analysis adopts the doping $\mu$ as the tunable parameter. This setup supports a tunable framework where the properties of the suppressed CDW phase can be examined.

\begin{figure}[b]
\centering
\includegraphics[scale = 1]{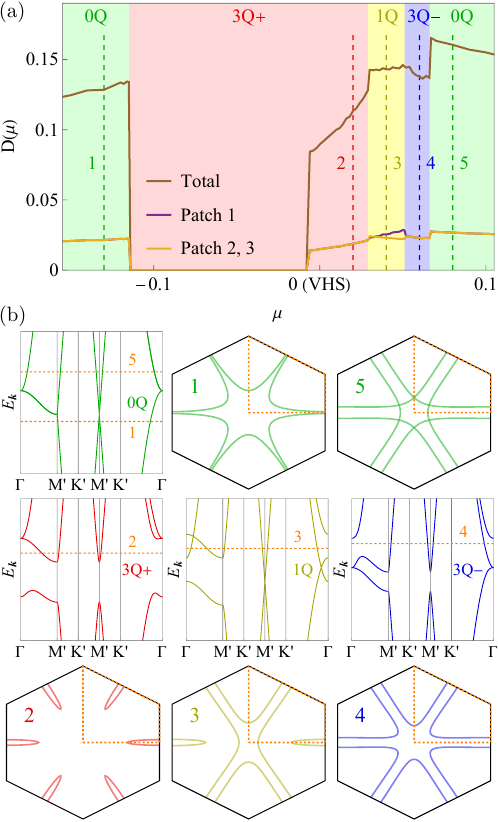}
\caption{\label{fig:rfs} Reconstructed Fermi surface under the parent CDW at $T=0$. We compute the density of states (a) and obtain the reconstructed band structure and Fermi surface (b). The density of states at each patch $D_\a$ is estimated by the total in each $\bsb\G$-$\mbf K'$-$\mbf K'$ $1/6$ subzone.}
\end{figure}

We are interested in the reconstructed Fermi surface under nonzero doping. To obtain the reconstructed band structure, we adopt the ground state of the effective Hamiltonian (\ref{eq:hcdw3band}) to a full four-band mean-field Hamiltonian in the reduced BZ \cite{lin21prb}. The choice of an $(s+d)$-wave form factor diminishes the CDW away from the Fermi level. Despite the mismatch in the model details, the computation captures the essential features of the reconstructed band structure. We monitor the evolution of the density of states $D(\mu)$ [Fig.~\ref{fig:rfs}(a)] and map out the reconstructed Fermi surface [Fig.~\ref{fig:rfs}(b)] at zero temperature $T=0$. In the $3Q+$ phase, the small $\mbf M'_\a$ pockets appear under the gaps at the Fermi level. The increasing doping enlarges the $\mbf M'_\a$ pockets and raises the density of states. Importantly, the enlarged $\mbf M'_\a$ pockets lose the energetic favor to the $\mbf K'_\a$ pockets at sufficiently large positive doping $\mu>0$. This corresponds to a transition to the $3Q-$ phase. A Lifshitz transition is naturally expected between the two phases, which hosts a reconstructed VHS. Although this transition is covered by an intermediate $1Q$ phase in our model, the proximity to the reconstructed VHS still enhances the density of states significantly. On the other hand, the reconstructed Fermi surface recovers its noninteracting form at the critical dopings. With the closeness to the VHS, the density of states is again elevated. Note that the reconstructed Fermi surface changes abruptly at the first-order transitions. Correspondingly, the density of states experiences sharp jumps at the regime boundaries.

Due to the presence of multiple transitions, the reconstructed metallic phase is divided into multiple regimes. Whether each regime appears is system dependent. Interestingly, the transitions between different CDW phases were observed under pressure \cite{du21prb,chen21prl,yu21nc,wang21prr}. Furthermore, the reconstructed VHS was probed close to the Fermi level for $A=\txt{Rb}$ \cite{cho21prl}. The multiregime nature of the reconstructed metal is essential to the low-energy phenomena.

\textit{Multidome superconductivity.---}As a central spirit of our work, we propose that the superconductivity develops as a weak-coupling instability on the reconstructed Fermi surface. This establishes a `parent-child relationship' between the CDW and the superconductivity, where the large separation of energy scales is naturally explained. Moreover, the multiregime reconstructed metal naturally hosts the intriguing multidome superconductivity. Deep in the CDW phase, the small density of states under the $3Q+$ orders supports the superconductivity at low critical temperature $T_\txt{SC}$. As the doping increases, the increasing density of states enhances the superconductivity and raises its critical temperature $T_\txt{SC}$. This is opposite to the suppression of the parent CDW phase, consistent with the experimental observations under pressure \cite{du21prb,chen21prl,yu21nc,wang21prr}, doping \cite{song21prl}, and uniaxial strain \cite{qian21prb}. Importantly, the reconstructed VHS can support significant enhancement and create a peak in the critical temperature $T_\txt{SC}$ \cite{nandkishore12np,lin19prb}. Although it is inaccessible in our model, the enhancement still manifests adequately across the $1Q$ regime. The elevated critical temperature $T_\txt{SC}$ depicts a superconducting dome in the parent CDW phase. When the CDW is driven across a critical doping, the superconductivity experiences another enhancement from the closeness to the VHS. This marks another superconducting dome which extends beyond the parent CDW phase. Thus our model predicts a `multidome' child superconductivity from the parent CDW (Fig.~\ref{fig:pd}). The double-dome structure under positive doping $\mu>0$ is close to the observations under pressure \cite{chen21prl,yu21nc,wang21prr}. Interestingly, the observed peaks are reminiscent of those from the (reconstructed) VHS. On the other hand, the double-dome structure under negative doping $\mu<0$ was also observed \cite{song21prl}, where the domes are separated by a sharp transition at the critical doping.

\textit{Patch model and symmetry analysis.---}To further understand the superconductivity, we adopt a low-energy patch model in the reconstructed metal. The patch model takes a six-patch form, where the small patches $\mbf p_\a$ capture the relevant portions of the reconstructed Fermi surface in the $\bsb\G$-$\mbf K'$-$\mbf K'$ $1/6$ subzones [Fig.~\ref{fig:setup}(c)]. The low-energy theory manifests the six-component fermion $(\psi_+^T,\psi_-^T)^T$ with $\psi_\pm=(\psi_{\pm1},\psi_{\pm2},\psi_{\pm3})^T$. Correspondingly, the density of states reads $(D^T,D^T)^T$ with $D=(D_1,D_2,D_3)^T$. The patch model offers a natural basis for the symmetry analysis. There exist six orthonormal representations $(w_+^T,w_-^T)^T$, which correspond to the form factors $w_{\mbf k}$ in six symmetry channels. Three of the channels are even and three are odd under the inversion $w_+=\pm w_-=w=w_{w_1w_2w_3}$, where $w_{w_1w_2w_3}$ denotes the normalized form of $(w_1,w_2,w_3)$. The symmetry channels are clearly illustrated by a mapping onto the Bloch sphere [Fig.~\ref{fig:setup}(c)]. At the ground state (GS) $w_\txt{GS}$, the characteristic orthonormal basis defines the channels as radial (RA) $w_\txt{RA}$, longitudinal (LO) $w_\txt{LO}$, and latitudinal (LA) $w_\txt{LA}$. Under the $\txt{C}_3$ symmetry, the ground state $w_\txt{GS}=w_{111}$ hosts $w_\txt{RA}=w_{111}$ along with the degenerate $w_\txt{LO}=w_{2-1-1}$ and $w_\txt{LA}=w_{01-1}$.

Note that the ground state may break the $\txt{C}_3$ symmetry, such as in the $1Q$ phase. Assume the $\txt{C}_3$ symmetry breaking at the momentum $\mbf Q_1$, which leads to a twofold symmetry along the $x$ direction. The reconstructed metal holds the (possibly weak) $x$ and $y$ reflection symmetries. In the patch model, the twofold symmetry manifests in the anisotropic patch momenta $\mbf p_1\neq\mbf p_2=\mbf p_3$ and density of states $D_1\neq D_2=D_3$. The ground state $w_\txt{GS}$ shifts away from $w_{111}$ along the longitude on the Bloch sphere [Fig.~\ref{fig:setup}(c)]. While the radial and longitudinal channels deform together in the reflection-even branch, the reflection-odd latitudinal channel remains invariant.

\textit{Pairing formalism.---}We now study the superconductivity in the low-energy theory. Consider the projection of general interaction on the Cooper channels \cite{lin18prb}
\beeq
H_\txt{CP}
=(P_\txt{CP}^\nu)^\dag V_{s/t}P_\txt{CP}^\nu.
\eneq
The spin-singlet and spin-triplet Cooper pairings (CPs) are $(P_\txt{CP}^\nu)^\dag_\a=\psi_{+\a}^\dag(\s^\nu/\sqrt2)[(i\s^2)(\psi_{-\a}^\dag)^T]$ with the spin Pauli matrices $\s^0=1$ and $\s^{1,2,3}$, respectively. Under the Fermi statistics, these pairings are coupled by the symmetric or antisymmetric interactions $V_{s\txt{/}t}=(V_{s\txt{/}t,\a\b})$ with $V_{s\txt{/}t,\a\b}=(V_{\mbf p_\a-\mbf p_\b}\pm V_{\mbf p_\a+\mbf p_\b})/2$. The interaction takes a general matrix form under the twofold symmetry
\beeq
V=\lf(\bear{ccc}V_1&V_3&V_3\\ V_3^*&V_2&V_4\\V_3^*&V_4&V_2\enar\ri).
\eneq
Here, $V_{1,2,4}\in\mbb R$ and $V_3\in\mbb C$ are assumed, which matches all candidates in our analysis. Note that $V_1=V_2$ and $V_3=V_4$ under the $\txt{C}_3$ symmetry.

A diagonalization determines the symmetry channels of the pairing states. In each symmetry channel
\beeq
H_{\txt{SC}}
=V_\txt{SC}(P_\txt{SC}^\nu)^\dag P_\txt{SC}^\nu,
\eneq
the pairing $(P_{\txt{SC}}^\nu)^\dag=\psi_+^\dag(\s^\nu/\sqrt2)\hat w_\txt{SC}[(i\s^2)(\psi_-^\dag)^T]$ adopts an eigenstate $w_\txt{SC}=\til w_\txt{SC}/|\til w_\txt{SC}|$ as the representation $\hat w_\txt{SC}=\txt{diag}(w_\txt{SC})$. The projected interaction $V_\txt{SC}$ is given by the corresponding eigenvalue. We identify the pairing states in six symmetry channels
\begin{flalign}
&\text{Radial or longitudinal}\hidewidth\nonumber\\
&& V_\txt{SC}=\fr{1}{2}(V_1+V_\pm),&\quad \til w_\txt{SC}=\lf(\bear{c}V_1-V_\mp\\2V_3^*\\2V_3^*\enar\ri), && \\
&\text{Latitudinal}\hidewidth\nonumber\\
&& V_\txt{SC}=V_2-V_4,& \quad \til w_\txt{SC}=\lf(\bear{c}0\\1\\-1\enar\ri), &&
\end{flalign}
with $V_\pm=V_2+V_4\pm\{[V_1-(V_2+V_4)]^2+8|V_3|^2\}^{1/2}$. Under the competition, the superconductivity develops from the leading pairing state with the strongest dimensionless interaction $\l_\txt{SC}=V_\txt{SC}(w_\txt{SC}^\dag Dw_\txt{SC})<0$ \cite{lin20prr}. The critical temperature is estimated as
\beeq
\label{eq:tc}
T_\txt{SC}=\L_\txt{SC}e^{-1/|\l_\txt{SC}|},
\eneq
where $\L_\txt{SC}$ is an energy cutoff in the low-energy theory.

A natural question arises as which pairing states are leading in the multidome superconductivity. Since the competition is interaction dependent, the answer is strongly contingent on the pairing mechanism. Here, we discuss some candidates and their resulting pairing states in the multidome superconductivity.

\textit{Phonon-mediated attraction.---}The phonon-mediated attraction always serves as a feasible candidate. Assume the weak short-range attraction with $V_{1,2,3,4}\apx V_0<0$. Under the $\txt{C}_3$ symmetry, the isotropic channel $w_{111}$ is leading. This leading channel becomes anisotropic when the twofold symmetry is manifest. Since the attraction is weakly momentum dependent, the anisotropy is weak. This contradicts with the strong anisotropy in the experimental observations \cite{ni21cpl,xiang21nc}. Therefore the phonon-mediated attraction may be an unlikely pairing mechanism.

\textit{CDW or RPOM fluctuation.---}The onset from the parent CDW suggests an intriguing pairing mechanism for the superconductivity. When the parent CDW manifests a certain fluctuation, the Cooper pairs can form through its mediation at low energy. This scenario naturally matches the large separation of energy scales between the superconductivity and the CDW. The parent CDW manifests itself in the reconstruction of the band structure. At low energy, this reconstruction turns into the RPOM order on the Fermi surface. Therefore the scenario is reminiscent of the pairing from the Pomeranchuk (POM) fluctuations \cite{fernandes13prl,fernandes14np,chen20prb}.

\begin{figure}[b]
\centering
\includegraphics[scale = 0.6]{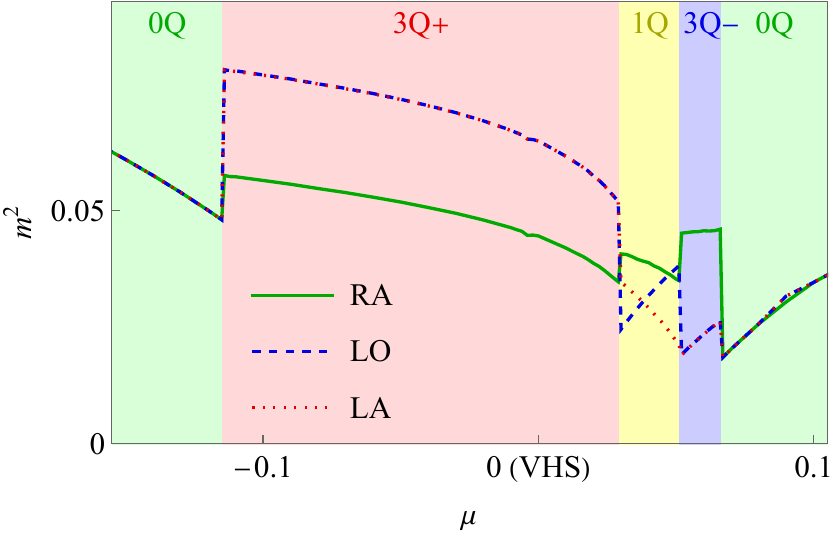}
\caption{\label{fig:mass} The masses of the CDW fluctuations at $T=0$.}
\end{figure}

We first discuss the fluctuations of the parent CDW. Diagonalizing the free-energy Hessian matrix $(\p_\a\p_\b f[\vec\D])_{\vec\D=\vec\D_\txt{GS}}$ at zero temperature $T=0$, we identify the symmetry channels of the CDW fluctuations (Fig.~\ref{fig:mass})
\beeq
\d f[\vec\D_\txt{GS}]=\fr{1}{2}m^2|\d\vec\D|^2
\eneq
with the representation $\d\vec\D=\d\D w$ and the effective mass $m^2$. Interestingly, these channels match the characteristic orthonormal basis $\{w_{\txt{RA},\txt{LO},\txt{LA}}\}$ at the ground state $\vec\D_\txt{GS}=\D w_\txt{GS}$ on the Bloch sphere [Fig.~\ref{fig:setup}(c)]. While the ground state exhibits the long-range CDW orders, the fluctuations exhibit the short-range correlations according to the nonzero masses.

In the parent CDW, the reconstructed metal acquires an inversion-even RPOM order $\vec\D_\txt{RPOM}=(\D_\txt{RPOM}/\sqrt2)(w^T,w^T)^T$. The RPOM order and its fluctuations correspond approximately to the CDW ones. Due to the positive semidefiniteness, the mapping onto the Bloch sphere now occurs in the first octant [Fig.~\ref{fig:setup}(c)]. The fluctuation in each symmetry channel is described by an effective charge density (CD) coupling \cite{abanov03advp,colemanmbp}
\beeq
H_{\d\D_\txt{CD}}=\sumv{q}\chi_{\mbf q}^{-1}\d\D_{\txt{CD},-\mbf q}\d\D_{\txt{CD},\mbf q}.
\eneq
The susceptibility $\chi_{\mbf q}>0$ is a result of tracing out the high-energy modes beyond the patch model. While the model involves the fluctuations at general momentum $\mbf q$, we assume that the susceptibility peaks nondivergently $\chi_{\mbf q=\mbf0}\apx2/m^2$ at zero momentum. The fluctuation couples to the CD pairing $P_{\txt{CD},\mbf q}=\sum_{\mbf k}\psi_{\mbf k+\mbf q}^\dag(\s^0/\sqrt2)w_{\mbf k,-\mbf q}\psi_{\mbf k}$ as
\beeq
H_{\psi\d\D_\txt{CD}}=g\sum_{\mbf q}P_{\txt{CD},\mbf q}\d\D_{\txt{CD},-\mbf q},
\eneq
where $w_{\mbf k,\mbf q}\sim\d\epvl{\psi_{\mbf k+\mbf q}^\dag\psi_{\mbf k}}$ is the form factor and $g$ is an effective coupling. Integrating out the fluctuation, we arrive at an effective attraction
\beeq
H_\txt{CD}=-g^2\sumv{q}\chi_{\mbf q}P_{\txt{CD},\mbf q}P_{\txt{CD},-\mbf q}.
\eneq

It is worth discussing the structure of the form factor $w_{\mbf k,\mbf q}$. The diagonal components $w_{\mbf p_\a,\mbf q=\mbf0}=\hat w_{\a\a}=w_\a$ correspond to the intrapatch RPOM representations. Meanwhile, the off-diagonal components $w_{\mbf p_\a,\mbf q=\mbf p_\b-\mbf p_\a}=\hat w_{\b\a}=\hat w_{\a\b}^*$ with $\a\neq\b$ are interpatch. Although the incommensurate momenta $\mbf p_\b-\mbf p_\a$ suggest the generally complex nature of the off-diagonal components $\hat w_{\b\a}$, the symmetries can fix some of the complex phases. While the inversion evenness implies $\hat w_{\b\a}=\hat w_{-\b-\a}$, the reflection evenness or oddness forces $\hat w_{\a\b}=\pm\hat  w_{\til\a\til\b}$ for the reflection pairs $(\a,\til \a)$ and $(\b,\til\b)$. In the $\txt{C}_3$ symmetric channel $w=w_{111}$, the off-diagonal components are $(\hat w_{2\pm3},\hat w_{3\pm1},\hat w_{1\pm2})=w_{111}$. On the other hand, the anisotropic channels may allow complex components due to the loss of certain reflection symmetries. Enforcing the reality, we find the components $(\hat w_{2\pm3},\hat w_{3\pm1},\hat w_{1\pm2})=w_{abb}$ with $a,b\in\mbb R$ and $w_{01-1}$ in the reflection-even and reflection-odd channels, respectively.

We now project the effective attraction on the Cooper channels
\beeq
H_\txt{CP}=-g^2\sum_{\a\b}\chi_{\a\b}\hat w_{\b\a}\hat w_{-\b-\a}\psi_\a^\dag\psi_{-\a}^\dag\psi_{-\b}\psi_\b.
\eneq
Here $\chi_{\a\b}=\chi_{\mbf q=\mbf p_\a-\mbf p_\b}$, and the four-fermion spin indices $(\s,\s',\s',\s)$ are suppressed. The symmetric or antisymmetric interactions read
\beeq
V_{\a\b}=-g^2(\chi_{\a\b}\hat w_{\b\a}^2\pm\chi_{\a-\b}\hat w_{-\b\a}^2).
\eneq
Since the symmetric interactions are generally stronger, the spin-singlet pairings are energetically favorable. The diagonal RPOM components determine the symmetry channels of the pairing states. Meanwhile, the off-diagonal components perturb these channels and break the possible degeneracy. The leading pairing state is chosen by the strongest dimensionless interaction
\beeq
\label{eq:dlint}
|\l_\txt{SC}|=C_\txt{SC}m^{-2}V_{w,\txt{SC}}(w_\txt{SC}^\dag Dw_\txt{SC}),
\eneq
which does not necessarily originate from the leading fluctuation. Here, $C_\txt{SC}$ is an effective parameter, and $V_{w,\txt{SC}}$ is an eigenvalue of the dominant interaction representation $\txt{diag}(w_1^2,w_2^2,w_3^2)$. Note that the CDW fluctuations can also trigger the superconductivity in the $0Q$ normal metal. With the unfolded patch model at $\mbf M_\a$, the symmetry channels are dominated by the off-diagonal CDW components.

Remarkably, the pairing states with the strongly anisotropic twofold structures $w_\txt{SC}=w_{1\d\d,\d11}$ can be leading in the multidome superconductivity (Table~\ref{tb:pairing}). Here, $0<\d\ll1$ (may be complex under the twofold symmetry) represents the infinitesimal components from the off-diagonal perturbations. Since the CDW breaks the $\txt{C}_3$ symmetry in the kagome metals $A\txt{V}_3\txt{Sb}_5$, the twofold structures are pinned along a single direction in the parent CDW phase. Interestingly, the twofold structures were observed in the magnetoresistance measurement \cite{ni21cpl,xiang21nc}. Furthermore, the sign preservation matches the scanning tunneling microscopy and spectroscopy \cite{xu21prl}. While a thermal conductivity measurement suggested a nodal gap \cite{zhao21axsc}, the measurements of magnetic penetration depth and specific heat caused an opposite conclusion to be drawn \cite{duan21scp}. This contradiction may originate from the infinitesimal component $\d$. Beyond the parent CDW phase, the degenerate pairing states $w_{011,101,110}$ may form the nodeless ground states with possible time-reversal symmetry breaking.

\begin{table}[t]
\centering
\betb{|c|c|c|}
\hline
GS&$w_\txt{GS}$&Fluctuation $\rar$ pairing state $w_\txt{SC}$\\
\hline
\multicolumn{3}{|c|}{In parent CDW phase}\\
\hline
$3Q+$&$w_{111}$&$w_\txt{LO}\rar w_{1\d\d}>w_\txt{LA}\rar\{w_{\d11}\gtrsim w_{01-1}\}$\\
\hline
$1Q$&$w_{100}$&$w_\txt{RA}\rar w_{1\d\d}>w_\txt{LO/LA}\rar\{w_{\d11}\gtrsim w_{01-1}\}$\\
\hline
$3Q-$&$w_{\mp111}$&$w_\txt{LO}\rar w_{1\d\d}>w_\txt{LA}\rar\{w_{\d11}\gtrsim w_{01-1}\}$\\
\hline
\multicolumn{3}{|c|}{Beyond parent CDW phase}\\
\hline
$0Q$&$w_{000}$&$w_\txt{RA/LO/LA}=w_{100,010,001}\rar w_{011,101,110}$\\
\hline
\entb
\caption{\label{tb:pairing} Leading pairing states from the CDW or RPOM fluctuations in the multidome superconductivity. Here, $w_1>w_2$ means that $w_1$ and $w_2$ are the leading and secondary channels, respectively, and $\{w_1\gtrsim w_2\}$ represents the degeneracy with perturbative breakdown. The $\mp$ sign in the $3Q-$ phase corresponds to the CDW or RPOM description.}
\end{table}

Finally, we note some features of the critical temperature $T_\txt{SC}$ from the CDW or RPOM fluctuations. The competition between different channels can broaden the transition temperature window between $T_\txt{SC}^\txt{onset}$ and $T_\txt{SC}^\txt{zero}$ \cite{chen21prl,yu21nc,wang21prr}, where the resistivity drops and vanishes, respectively. Meanwhile, the enhanced fluctuations at the first-order transitions can strengthen the peaks of the critical temperature $T_\txt{SC}$ at the regime boundaries.


\textit{Discussion.---}We show that the parent CDW naturally hosts the multidome child superconductivity in the kagome metals $A\txt{V}_3\txt{Sb}_5$. The `parent-child relationship' realizes the large separation of energy scales between the superconductivity and the CDW. Meanwhile, the multidome superconductivity originates from the multiregime reconstructed metal with distinct (reconstructed) VHSs. The pairing states with strong twofold anisotropy can develop from the CDW or RPOM fluctuations. Our work sheds light on an unconventional pairing mechanism with strong evidence in the kagome metals $A\txt{V}_3\txt{Sb}_5$.

Our analysis is exemplified with a real CDW on a two-dimensional single-orbital kagome lattice. The inclusion of three-dimensional multiband orders may yield a more accurate description of the kagome metals $A\txt{V}_3\txt{Sb}_5$, such as the general involvement of twofold symmetry \cite{christensen21prb} and the access to the reconstructed VHS. A combination with an imaginary CDW \cite{lin21prb} can further incorporate the time-reversal symmetry breaking. Meanwhile, the possible band topology \cite{yang20sa,yu21prb,ortiz20prl,ortiz21prm,lin21prb,lin21ax} can make the superconductivity geometrically enhanced \cite{peotta15nc,liang17prb,hu19prl,xie20prl,lin21prbdh} or topological \cite{li18prl}. Spin-orbit coupling may also lead to additional features. Note that the multidome structure follows solely from the evolution of the density of states, which will be shared by any weak-coupling pairing mechanism on the reconstructed Fermi surface. If the ferromagnetic fluctuation or the Kohn-Luttinger renormalization \cite{nandkishore14prb,lin18prb} is strong, the spin-triplet pairing states may develop. On the other hand, the first-order transitions may hinder the quantum critical behavior at the regime boundaries \cite{berg12s,rodriguez21prb}. The related discussion is beyond our mean-field framework and is an interesting topic for future work. Finally, since the VH Fermi surface is universal on the hexagonal lattices, our analysis is also applicable to the other hexagonal-lattice systems.

{\it Note added.} Recently, we learned about an independent study of kagome superconductors from CDW fluctuations \cite{tazai2sa}. This work considered the unfolded unreconstructed theory, which is eligible outside the CDW. We adopt the folded reconstructed theory in the parent CDW phase. This captures more precisely the situation in the kagome metals $A\txt{V}_3\txt{Sb}_5$ and explains the double-dome superconductivity. Furthermore, our work conducts a systematic symmetry analysis of the multi-$Q$ CDW orders and their fluctuations. This determines the leading pairing states with strong experimental relevance in a complete framework.

\begin{acknowledgments}
This research was sponsored by the Army Research Office and was accomplished under Grant No. W911NF-17-1-0482. The views and conclusions contained in this document are those of the authors and should not be interpreted as representing the official policies, either expressed or implied, of the Army Research Office or the U.S. Government. The U.S. Government is authorized to reproduce and distribute reprints for Government purposes notwithstanding any copyright notation herein. RN also acknowledges the support of the Alfred P. Sloan foundation through a Sloan Research Fellowship. 
\end{acknowledgments}




\bibliography{Reference}

\begin{thebibliography}{80}%
\makeatletter
\providecommand \@ifxundefined [1]{%
 \@ifx{#1\undefined}
}%
\providecommand \@ifnum [1]{%
 \ifnum #1\expandafter \@firstoftwo
 \else \expandafter \@secondoftwo
 \fi
}%
\providecommand \@ifx [1]{%
 \ifx #1\expandafter \@firstoftwo
 \else \expandafter \@secondoftwo
 \fi
}%
\providecommand \natexlab [1]{#1}%
\providecommand \enquote  [1]{``#1''}%
\providecommand \bibnamefont  [1]{#1}%
\providecommand \bibfnamefont [1]{#1}%
\providecommand \citenamefont [1]{#1}%
\providecommand \href@noop [0]{\@secondoftwo}%
\providecommand \href [0]{\begingroup \@sanitize@url \@href}%
\providecommand \@href[1]{\@@startlink{#1}\@@href}%
\providecommand \@@href[1]{\endgroup#1\@@endlink}%
\providecommand \@sanitize@url [0]{\catcode `\\12\catcode `\$12\catcode
  `\&12\catcode `\#12\catcode `\^12\catcode `\_12\catcode `\%12\relax}%
\providecommand \@@startlink[1]{}%
\providecommand \@@endlink[0]{}%
\providecommand \url  [0]{\begingroup\@sanitize@url \@url }%
\providecommand \@url [1]{\endgroup\@href {#1}{\urlprefix }}%
\providecommand \urlprefix  [0]{URL }%
\providecommand \Eprint [0]{\href }%
\providecommand \doibase [0]{https://doi.org/}%
\providecommand \selectlanguage [0]{\@gobble}%
\providecommand \bibinfo  [0]{\@secondoftwo}%
\providecommand \bibfield  [0]{\@secondoftwo}%
\providecommand \translation [1]{[#1]}%
\providecommand \BibitemOpen [0]{}%
\providecommand \bibitemStop [0]{}%
\providecommand \bibitemNoStop [0]{.\EOS\space}%
\providecommand \EOS [0]{\spacefactor3000\relax}%
\providecommand \BibitemShut  [1]{\csname bibitem#1\endcsname}%
\let\auto@bib@innerbib\@empty
\bibitem [{\citenamefont {Ortiz}\ \emph {et~al.}(2019)\citenamefont {Ortiz},
  \citenamefont {Gomes}, \citenamefont {Morey}, \citenamefont {Winiarski},
  \citenamefont {Bordelon}, \citenamefont {Mangum}, \citenamefont {Oswald},
  \citenamefont {Rodriguez-Rivera}, \citenamefont {Neilson}, \citenamefont
  {Wilson}, \citenamefont {Ertekin}, \citenamefont {McQueen},\ and\
  \citenamefont {Toberer}}]{ortiz19prm}%
  \BibitemOpen
  \bibfield  {author} {\bibinfo {author} {\bibfnamefont {B.~R.}\ \bibnamefont
  {Ortiz}}, \bibinfo {author} {\bibfnamefont {L.~C.}\ \bibnamefont {Gomes}},
  \bibinfo {author} {\bibfnamefont {J.~R.}\ \bibnamefont {Morey}}, \bibinfo
  {author} {\bibfnamefont {M.}~\bibnamefont {Winiarski}}, \bibinfo {author}
  {\bibfnamefont {M.}~\bibnamefont {Bordelon}}, \bibinfo {author}
  {\bibfnamefont {J.~S.}\ \bibnamefont {Mangum}}, \bibinfo {author}
  {\bibfnamefont {I.~W.~H.}\ \bibnamefont {Oswald}}, \bibinfo {author}
  {\bibfnamefont {J.~A.}\ \bibnamefont {Rodriguez-Rivera}}, \bibinfo {author}
  {\bibfnamefont {J.~R.}\ \bibnamefont {Neilson}}, \bibinfo {author}
  {\bibfnamefont {S.~D.}\ \bibnamefont {Wilson}}, \bibinfo {author}
  {\bibfnamefont {E.}~\bibnamefont {Ertekin}}, \bibinfo {author} {\bibfnamefont
  {T.~M.}\ \bibnamefont {McQueen}},\ and\ \bibinfo {author} {\bibfnamefont
  {E.~S.}\ \bibnamefont {Toberer}},\ }\bibfield  {title} {\bibinfo {title}
  {{New kagome prototype materials: discovery of
  ${\mathrm{KV}}_{3}{\mathrm{Sb}}_{5},{\mathrm{RbV}}_{3}{\mathrm{Sb}}_{5}$, and
  ${\mathrm{CsV}}_{3}{\mathrm{Sb}}_{5}$}},\ }\href
  {https://doi.org/10.1103/PhysRevMaterials.3.094407} {\bibfield  {journal}
  {\bibinfo  {journal} {Phys. Rev. Materials}\ }\textbf {\bibinfo {volume}
  {3}},\ \bibinfo {pages} {094407} (\bibinfo {year} {2019})}\BibitemShut
  {NoStop}%
\bibitem [{\citenamefont {Yang}\ \emph {et~al.}(2020)\citenamefont {Yang},
  \citenamefont {Wang}, \citenamefont {Ortiz}, \citenamefont {Liu},
  \citenamefont {Gayles}, \citenamefont {Derunova}, \citenamefont
  {Gonzalez-Hernandez}, \citenamefont {{\v S}mejkal}, \citenamefont {Chen},
  \citenamefont {Parkin}, \citenamefont {Wilson}, \citenamefont {Toberer},
  \citenamefont {McQueen},\ and\ \citenamefont {Ali}}]{yang20sa}%
  \BibitemOpen
  \bibfield  {author} {\bibinfo {author} {\bibfnamefont {S.-Y.}\ \bibnamefont
  {Yang}}, \bibinfo {author} {\bibfnamefont {Y.}~\bibnamefont {Wang}}, \bibinfo
  {author} {\bibfnamefont {B.~R.}\ \bibnamefont {Ortiz}}, \bibinfo {author}
  {\bibfnamefont {D.}~\bibnamefont {Liu}}, \bibinfo {author} {\bibfnamefont
  {J.}~\bibnamefont {Gayles}}, \bibinfo {author} {\bibfnamefont
  {E.}~\bibnamefont {Derunova}}, \bibinfo {author} {\bibfnamefont
  {R.}~\bibnamefont {Gonzalez-Hernandez}}, \bibinfo {author} {\bibfnamefont
  {L.}~\bibnamefont {{\v S}mejkal}}, \bibinfo {author} {\bibfnamefont
  {Y.}~\bibnamefont {Chen}}, \bibinfo {author} {\bibfnamefont {S.~S.~P.}\
  \bibnamefont {Parkin}}, \bibinfo {author} {\bibfnamefont {S.~D.}\
  \bibnamefont {Wilson}}, \bibinfo {author} {\bibfnamefont {E.~S.}\
  \bibnamefont {Toberer}}, \bibinfo {author} {\bibfnamefont {T.}~\bibnamefont
  {McQueen}},\ and\ \bibinfo {author} {\bibfnamefont {M.~N.}\ \bibnamefont
  {Ali}},\ }\bibfield  {title} {\bibinfo {title} {{Giant, unconventional
  anomalous Hall effect in the metallic frustrated magnet candidate,
  KV$_3$Sb$_5$}},\ }\href {https://doi.org/10.1126/sciadv.abb6003} {\bibfield
  {journal} {\bibinfo  {journal} {Sci. Adv.}\ }\textbf {\bibinfo {volume}
  {6}},\ \bibinfo {pages} {eabb6003} (\bibinfo {year} {2020})}\BibitemShut
  {NoStop}%
\bibitem [{\citenamefont {Ortiz}\ \emph {et~al.}(2020)\citenamefont {Ortiz},
  \citenamefont {Teicher}, \citenamefont {Hu}, \citenamefont {Zuo},
  \citenamefont {Sarte}, \citenamefont {Schueller}, \citenamefont {Abeykoon},
  \citenamefont {Krogstad}, \citenamefont {Rosenkranz}, \citenamefont {Osborn},
  \citenamefont {Seshadri}, \citenamefont {Balents}, \citenamefont {He},\ and\
  \citenamefont {Wilson}}]{ortiz20prl}%
  \BibitemOpen
  \bibfield  {author} {\bibinfo {author} {\bibfnamefont {B.~R.}\ \bibnamefont
  {Ortiz}}, \bibinfo {author} {\bibfnamefont {S.~M.~L.}\ \bibnamefont
  {Teicher}}, \bibinfo {author} {\bibfnamefont {Y.}~\bibnamefont {Hu}},
  \bibinfo {author} {\bibfnamefont {J.~L.}\ \bibnamefont {Zuo}}, \bibinfo
  {author} {\bibfnamefont {P.~M.}\ \bibnamefont {Sarte}}, \bibinfo {author}
  {\bibfnamefont {E.~C.}\ \bibnamefont {Schueller}}, \bibinfo {author}
  {\bibfnamefont {A.~M.~M.}\ \bibnamefont {Abeykoon}}, \bibinfo {author}
  {\bibfnamefont {M.~J.}\ \bibnamefont {Krogstad}}, \bibinfo {author}
  {\bibfnamefont {S.}~\bibnamefont {Rosenkranz}}, \bibinfo {author}
  {\bibfnamefont {R.}~\bibnamefont {Osborn}}, \bibinfo {author} {\bibfnamefont
  {R.}~\bibnamefont {Seshadri}}, \bibinfo {author} {\bibfnamefont
  {L.}~\bibnamefont {Balents}}, \bibinfo {author} {\bibfnamefont
  {J.}~\bibnamefont {He}},\ and\ \bibinfo {author} {\bibfnamefont {S.~D.}\
  \bibnamefont {Wilson}},\ }\bibfield  {title} {\bibinfo {title}
  {{$\mathrm{Cs}{\mathrm{V}}_{3}{\mathrm{Sb}}_{5}$: A ${\mathbb{Z}}_{2}$
  Topological Kagome Metal with a Superconducting Ground State}},\ }\href
  {https://doi.org/10.1103/PhysRevLett.125.247002} {\bibfield  {journal}
  {\bibinfo  {journal} {Phys. Rev. Lett.}\ }\textbf {\bibinfo {volume} {125}},\
  \bibinfo {pages} {247002} (\bibinfo {year} {2020})}\BibitemShut {NoStop}%
\bibitem [{\citenamefont {Kenney}\ \emph {et~al.}(2021)\citenamefont {Kenney},
  \citenamefont {Ortiz}, \citenamefont {Wang}, \citenamefont {Wilson},\ and\
  \citenamefont {Graf}}]{kenney21jpcm}%
  \BibitemOpen
  \bibfield  {author} {\bibinfo {author} {\bibfnamefont {E.~M.}\ \bibnamefont
  {Kenney}}, \bibinfo {author} {\bibfnamefont {B.~R.}\ \bibnamefont {Ortiz}},
  \bibinfo {author} {\bibfnamefont {C.}~\bibnamefont {Wang}}, \bibinfo {author}
  {\bibfnamefont {S.~D.}\ \bibnamefont {Wilson}},\ and\ \bibinfo {author}
  {\bibfnamefont {M.~J.}\ \bibnamefont {Graf}},\ }\bibfield  {title} {\bibinfo
  {title} {{Absence of local moments in the kagome metal {KV}$_3$Sb$_5$ as
  determined by muon spin spectroscopy}},\ }\href
  {https://doi.org/10.1088/1361-648x/abe8f9} {\bibfield  {journal} {\bibinfo
  {journal} {J. Phys.: Condens. Matter}\ }\textbf {\bibinfo {volume} {33}},\
  \bibinfo {pages} {235801} (\bibinfo {year} {2021})}\BibitemShut {NoStop}%
\bibitem [{\citenamefont {{Jiang}}\ \emph {et~al.}(2021)\citenamefont
  {{Jiang}}, \citenamefont {{Yin}}, \citenamefont {{Denner}}, \citenamefont
  {{Shumiya}}, \citenamefont {{Ortiz}}, \citenamefont {{Xu}}, \citenamefont
  {{Guguchia}}, \citenamefont {{He}}, \citenamefont {{Hossain}}, \citenamefont
  {{Liu}}, \citenamefont {{Ruff}}, \citenamefont {{Kautzsch}}, \citenamefont
  {{Zhang}}, \citenamefont {{Chang}}, \citenamefont {{Belopolski}},
  \citenamefont {{Zhang}}, \citenamefont {{Cochran}}, \citenamefont {{Multer}},
  \citenamefont {{Litskevich}}, \citenamefont {{Cheng}}, \citenamefont
  {{Yang}}, \citenamefont {{Wang}}, \citenamefont {{Thomale}}, \citenamefont
  {{Neupert}}, \citenamefont {{Wilson}},\ and\ \citenamefont {{Zahid
  Hasan}}}]{jiang21nm}%
  \BibitemOpen
  \bibfield  {author} {\bibinfo {author} {\bibfnamefont {Y.-X.}\ \bibnamefont
  {{Jiang}}}, \bibinfo {author} {\bibfnamefont {J.-X.}\ \bibnamefont {{Yin}}},
  \bibinfo {author} {\bibfnamefont {M.~M.}\ \bibnamefont {{Denner}}}, \bibinfo
  {author} {\bibfnamefont {N.}~\bibnamefont {{Shumiya}}}, \bibinfo {author}
  {\bibfnamefont {B.~R.}\ \bibnamefont {{Ortiz}}}, \bibinfo {author}
  {\bibfnamefont {G.}~\bibnamefont {{Xu}}}, \bibinfo {author} {\bibfnamefont
  {Z.}~\bibnamefont {{Guguchia}}}, \bibinfo {author} {\bibfnamefont
  {J.}~\bibnamefont {{He}}}, \bibinfo {author} {\bibfnamefont {M.~S.}\
  \bibnamefont {{Hossain}}}, \bibinfo {author} {\bibfnamefont {X.}~\bibnamefont
  {{Liu}}}, \bibinfo {author} {\bibfnamefont {J.}~\bibnamefont {{Ruff}}},
  \bibinfo {author} {\bibfnamefont {L.}~\bibnamefont {{Kautzsch}}}, \bibinfo
  {author} {\bibfnamefont {S.~S.}\ \bibnamefont {{Zhang}}}, \bibinfo {author}
  {\bibfnamefont {G.}~\bibnamefont {{Chang}}}, \bibinfo {author} {\bibfnamefont
  {I.}~\bibnamefont {{Belopolski}}}, \bibinfo {author} {\bibfnamefont
  {Q.}~\bibnamefont {{Zhang}}}, \bibinfo {author} {\bibfnamefont {T.~A.}\
  \bibnamefont {{Cochran}}}, \bibinfo {author} {\bibfnamefont {D.}~\bibnamefont
  {{Multer}}}, \bibinfo {author} {\bibfnamefont {M.}~\bibnamefont
  {{Litskevich}}}, \bibinfo {author} {\bibfnamefont {Z.-J.}\ \bibnamefont
  {{Cheng}}}, \bibinfo {author} {\bibfnamefont {X.~P.}\ \bibnamefont {{Yang}}},
  \bibinfo {author} {\bibfnamefont {Z.}~\bibnamefont {{Wang}}}, \bibinfo
  {author} {\bibfnamefont {R.}~\bibnamefont {{Thomale}}}, \bibinfo {author}
  {\bibfnamefont {T.}~\bibnamefont {{Neupert}}}, \bibinfo {author}
  {\bibfnamefont {S.~D.}\ \bibnamefont {{Wilson}}},\ and\ \bibinfo {author}
  {\bibfnamefont {M.}~\bibnamefont {{Zahid Hasan}}},\ }\bibfield  {title}
  {\bibinfo {title} {{Unconventional chiral charge order in kagome
  superconductor KV$_3$Sb$_5$}},\ }\href
  {https://doi.org/10.1038/s41563-021-01034-y} {\bibfield  {journal} {\bibinfo
  {journal} {Nat. Mater.}\ }\textbf {\bibinfo {volume} {20}},\ \bibinfo {pages}
  {1353} (\bibinfo {year} {2021})}\BibitemShut {NoStop}%
\bibitem [{\citenamefont {Chen}\ \emph {et~al.}(2021)\citenamefont {Chen},
  \citenamefont {Wang}, \citenamefont {Yin}, \citenamefont {Gu}, \citenamefont
  {Jiang}, \citenamefont {Tu}, \citenamefont {Gong}, \citenamefont {Uwatoko},
  \citenamefont {Sun}, \citenamefont {Lei}, \citenamefont {Hu},\ and\
  \citenamefont {Cheng}}]{chen21prl}%
  \BibitemOpen
  \bibfield  {author} {\bibinfo {author} {\bibfnamefont {K.~Y.}\ \bibnamefont
  {Chen}}, \bibinfo {author} {\bibfnamefont {N.~N.}\ \bibnamefont {Wang}},
  \bibinfo {author} {\bibfnamefont {Q.~W.}\ \bibnamefont {Yin}}, \bibinfo
  {author} {\bibfnamefont {Y.~H.}\ \bibnamefont {Gu}}, \bibinfo {author}
  {\bibfnamefont {K.}~\bibnamefont {Jiang}}, \bibinfo {author} {\bibfnamefont
  {Z.~J.}\ \bibnamefont {Tu}}, \bibinfo {author} {\bibfnamefont {C.~S.}\
  \bibnamefont {Gong}}, \bibinfo {author} {\bibfnamefont {Y.}~\bibnamefont
  {Uwatoko}}, \bibinfo {author} {\bibfnamefont {J.~P.}\ \bibnamefont {Sun}},
  \bibinfo {author} {\bibfnamefont {H.~C.}\ \bibnamefont {Lei}}, \bibinfo
  {author} {\bibfnamefont {J.~P.}\ \bibnamefont {Hu}},\ and\ \bibinfo {author}
  {\bibfnamefont {J.-G.}\ \bibnamefont {Cheng}},\ }\bibfield  {title} {\bibinfo
  {title} {{Double Superconducting Dome and Triple Enhancement of ${T}_{c}$ in
  the Kagome Superconductor ${\mathrm{CsV}}_{3}{\mathrm{Sb}}_{5}$ under High
  Pressure}},\ }\href {https://doi.org/10.1103/PhysRevLett.126.247001}
  {\bibfield  {journal} {\bibinfo  {journal} {Phys. Rev. Lett.}\ }\textbf
  {\bibinfo {volume} {126}},\ \bibinfo {pages} {247001} (\bibinfo {year}
  {2021})}\BibitemShut {NoStop}%
\bibitem [{\citenamefont {Yu}\ \emph {et~al.}(2021{\natexlab{a}})\citenamefont
  {Yu}, \citenamefont {Wu}, \citenamefont {Wang}, \citenamefont {Lei},
  \citenamefont {Zhuo}, \citenamefont {Ying},\ and\ \citenamefont
  {Chen}}]{yu21prb}%
  \BibitemOpen
  \bibfield  {author} {\bibinfo {author} {\bibfnamefont {F.~H.}\ \bibnamefont
  {Yu}}, \bibinfo {author} {\bibfnamefont {T.}~\bibnamefont {Wu}}, \bibinfo
  {author} {\bibfnamefont {Z.~Y.}\ \bibnamefont {Wang}}, \bibinfo {author}
  {\bibfnamefont {B.}~\bibnamefont {Lei}}, \bibinfo {author} {\bibfnamefont
  {W.~Z.}\ \bibnamefont {Zhuo}}, \bibinfo {author} {\bibfnamefont {J.~J.}\
  \bibnamefont {Ying}},\ and\ \bibinfo {author} {\bibfnamefont {X.~H.}\
  \bibnamefont {Chen}},\ }\bibfield  {title} {\bibinfo {title} {Concurrence of
  anomalous hall effect and charge density wave in a superconducting
  topological kagome metal},\ }\href
  {https://doi.org/10.1103/PhysRevB.104.L041103} {\bibfield  {journal}
  {\bibinfo  {journal} {Phys. Rev. B}\ }\textbf {\bibinfo {volume} {104}},\
  \bibinfo {pages} {L041103} (\bibinfo {year}
  {2021}{\natexlab{a}})}\BibitemShut {NoStop}%
\bibitem [{\citenamefont {Du}\ \emph {et~al.}(2021)\citenamefont {Du},
  \citenamefont {Luo}, \citenamefont {Ortiz}, \citenamefont {Chen},
  \citenamefont {Duan}, \citenamefont {Zhang}, \citenamefont {Lu},
  \citenamefont {Wilson}, \citenamefont {Song},\ and\ \citenamefont
  {Yuan}}]{du21prb}%
  \BibitemOpen
  \bibfield  {author} {\bibinfo {author} {\bibfnamefont {F.}~\bibnamefont
  {Du}}, \bibinfo {author} {\bibfnamefont {S.}~\bibnamefont {Luo}}, \bibinfo
  {author} {\bibfnamefont {B.~R.}\ \bibnamefont {Ortiz}}, \bibinfo {author}
  {\bibfnamefont {Y.}~\bibnamefont {Chen}}, \bibinfo {author} {\bibfnamefont
  {W.}~\bibnamefont {Duan}}, \bibinfo {author} {\bibfnamefont {D.}~\bibnamefont
  {Zhang}}, \bibinfo {author} {\bibfnamefont {X.}~\bibnamefont {Lu}}, \bibinfo
  {author} {\bibfnamefont {S.~D.}\ \bibnamefont {Wilson}}, \bibinfo {author}
  {\bibfnamefont {Y.}~\bibnamefont {Song}},\ and\ \bibinfo {author}
  {\bibfnamefont {H.}~\bibnamefont {Yuan}},\ }\bibfield  {title} {\bibinfo
  {title} {{Pressure-induced double superconducting domes and charge
  instability in the kagome metal ${\mathrm{KV}}_{3}{\mathrm{Sb}}_{5}$}},\
  }\href {https://doi.org/10.1103/PhysRevB.103.L220504} {\bibfield  {journal}
  {\bibinfo  {journal} {Phys. Rev. B}\ }\textbf {\bibinfo {volume} {103}},\
  \bibinfo {pages} {L220504} (\bibinfo {year} {2021})}\BibitemShut {NoStop}%
\bibitem [{\citenamefont {{Zhao}}\ \emph
  {et~al.}(2021{\natexlab{a}})\citenamefont {{Zhao}}, \citenamefont {{Li}},
  \citenamefont {{Ortiz}}, \citenamefont {{Teicher}}, \citenamefont {{Park}},
  \citenamefont {{Ye}}, \citenamefont {{Wang}}, \citenamefont {{Balents}},
  \citenamefont {{Wilson}},\ and\ \citenamefont {{Zeljkovic}}}]{zhao21n}%
  \BibitemOpen
  \bibfield  {author} {\bibinfo {author} {\bibfnamefont {H.}~\bibnamefont
  {{Zhao}}}, \bibinfo {author} {\bibfnamefont {H.}~\bibnamefont {{Li}}},
  \bibinfo {author} {\bibfnamefont {B.~R.}\ \bibnamefont {{Ortiz}}}, \bibinfo
  {author} {\bibfnamefont {S.~M.~L.}\ \bibnamefont {{Teicher}}}, \bibinfo
  {author} {\bibfnamefont {T.}~\bibnamefont {{Park}}}, \bibinfo {author}
  {\bibfnamefont {M.}~\bibnamefont {{Ye}}}, \bibinfo {author} {\bibfnamefont
  {Z.}~\bibnamefont {{Wang}}}, \bibinfo {author} {\bibfnamefont
  {L.}~\bibnamefont {{Balents}}}, \bibinfo {author} {\bibfnamefont {S.~D.}\
  \bibnamefont {{Wilson}}},\ and\ \bibinfo {author} {\bibfnamefont
  {I.}~\bibnamefont {{Zeljkovic}}},\ }\bibfield  {title} {\bibinfo {title}
  {{Cascade of correlated electron states in a kagome superconductor
  CsV$_3$Sb$_5$}},\ }\href {https://doi.org/10.1038/s41586-021-03946-w}
  {\bibfield  {journal} {\bibinfo  {journal} {Nature}\ }\textbf {\bibinfo
  {volume} {599}},\ \bibinfo {pages} {216} (\bibinfo {year}
  {2021}{\natexlab{a}})}\BibitemShut {NoStop}%
\bibitem [{\citenamefont {Liang}\ \emph {et~al.}(2021)\citenamefont {Liang},
  \citenamefont {Hou}, \citenamefont {Zhang}, \citenamefont {Ma}, \citenamefont
  {Wu}, \citenamefont {Zhang}, \citenamefont {Yu}, \citenamefont {Ying},
  \citenamefont {Jiang}, \citenamefont {Shan}, \citenamefont {Wang},\ and\
  \citenamefont {Chen}}]{liang21prx}%
  \BibitemOpen
  \bibfield  {author} {\bibinfo {author} {\bibfnamefont {Z.}~\bibnamefont
  {Liang}}, \bibinfo {author} {\bibfnamefont {X.}~\bibnamefont {Hou}}, \bibinfo
  {author} {\bibfnamefont {F.}~\bibnamefont {Zhang}}, \bibinfo {author}
  {\bibfnamefont {W.}~\bibnamefont {Ma}}, \bibinfo {author} {\bibfnamefont
  {P.}~\bibnamefont {Wu}}, \bibinfo {author} {\bibfnamefont {Z.}~\bibnamefont
  {Zhang}}, \bibinfo {author} {\bibfnamefont {F.}~\bibnamefont {Yu}}, \bibinfo
  {author} {\bibfnamefont {J.-J.}\ \bibnamefont {Ying}}, \bibinfo {author}
  {\bibfnamefont {K.}~\bibnamefont {Jiang}}, \bibinfo {author} {\bibfnamefont
  {L.}~\bibnamefont {Shan}}, \bibinfo {author} {\bibfnamefont {Z.}~\bibnamefont
  {Wang}},\ and\ \bibinfo {author} {\bibfnamefont {X.-H.}\ \bibnamefont
  {Chen}},\ }\bibfield  {title} {\bibinfo {title} {{Three-Dimensional Charge
  Density Wave and Surface-Dependent Vortex-Core States in a Kagome
  Superconductor ${\mathrm{CsV}}_{3}{\mathrm{Sb}}_{5}$}},\ }\href
  {https://doi.org/10.1103/PhysRevX.11.031026} {\bibfield  {journal} {\bibinfo
  {journal} {Phys. Rev. X}\ }\textbf {\bibinfo {volume} {11}},\ \bibinfo
  {pages} {031026} (\bibinfo {year} {2021})}\BibitemShut {NoStop}%
\bibitem [{\citenamefont {{Uykur}}\ \emph {et~al.}(2022)\citenamefont
  {{Uykur}}, \citenamefont {{Ortiz}}, \citenamefont {{Wilson}}, \citenamefont
  {{Dressel}},\ and\ \citenamefont {{Tsirlin}}}]{uykur22npjqm}%
  \BibitemOpen
  \bibfield  {author} {\bibinfo {author} {\bibfnamefont {E.}~\bibnamefont
  {{Uykur}}}, \bibinfo {author} {\bibfnamefont {B.~R.}\ \bibnamefont
  {{Ortiz}}}, \bibinfo {author} {\bibfnamefont {S.~D.}\ \bibnamefont
  {{Wilson}}}, \bibinfo {author} {\bibfnamefont {M.}~\bibnamefont
  {{Dressel}}},\ and\ \bibinfo {author} {\bibfnamefont {A.~A.}\ \bibnamefont
  {{Tsirlin}}},\ }\bibfield  {title} {\bibinfo {title} {{Optical detection of
  the density-wave instability in the kagome metal KV$_3$Sb$_5$}},\ }\href
  {https://doi.org/10.1038/s41535-021-00420-8} {\bibfield  {journal} {\bibinfo
  {journal} {npj Quantum Mater.}\ }\textbf {\bibinfo {volume} {7}},\ \bibinfo
  {pages} {16} (\bibinfo {year} {2022})}\BibitemShut {NoStop}%
\bibitem [{\citenamefont {{Chen}}\ \emph {et~al.}(2021)\citenamefont {{Chen}},
  \citenamefont {{Yang}}, \citenamefont {{Hu}}, \citenamefont {{Zhao}},
  \citenamefont {{Yuan}}, \citenamefont {{Xing}}, \citenamefont {{Qian}},
  \citenamefont {{Huang}}, \citenamefont {{Li}}, \citenamefont {{Ye}},
  \citenamefont {{Ma}}, \citenamefont {{Ni}}, \citenamefont {{Zhang}},
  \citenamefont {{Yin}}, \citenamefont {{Gong}}, \citenamefont {{Tu}},
  \citenamefont {{Lei}}, \citenamefont {{Tan}}, \citenamefont {{Zhou}},
  \citenamefont {{Shen}}, \citenamefont {{Dong}}, \citenamefont {{Yan}},
  \citenamefont {{Wang}},\ and\ \citenamefont {{Gao}}}]{chen21n}%
  \BibitemOpen
  \bibfield  {author} {\bibinfo {author} {\bibfnamefont {H.}~\bibnamefont
  {{Chen}}}, \bibinfo {author} {\bibfnamefont {H.}~\bibnamefont {{Yang}}},
  \bibinfo {author} {\bibfnamefont {B.}~\bibnamefont {{Hu}}}, \bibinfo {author}
  {\bibfnamefont {Z.}~\bibnamefont {{Zhao}}}, \bibinfo {author} {\bibfnamefont
  {J.}~\bibnamefont {{Yuan}}}, \bibinfo {author} {\bibfnamefont
  {Y.}~\bibnamefont {{Xing}}}, \bibinfo {author} {\bibfnamefont
  {G.}~\bibnamefont {{Qian}}}, \bibinfo {author} {\bibfnamefont
  {Z.}~\bibnamefont {{Huang}}}, \bibinfo {author} {\bibfnamefont
  {G.}~\bibnamefont {{Li}}}, \bibinfo {author} {\bibfnamefont {Y.}~\bibnamefont
  {{Ye}}}, \bibinfo {author} {\bibfnamefont {S.}~\bibnamefont {{Ma}}}, \bibinfo
  {author} {\bibfnamefont {S.}~\bibnamefont {{Ni}}}, \bibinfo {author}
  {\bibfnamefont {H.}~\bibnamefont {{Zhang}}}, \bibinfo {author} {\bibfnamefont
  {Q.}~\bibnamefont {{Yin}}}, \bibinfo {author} {\bibfnamefont
  {C.}~\bibnamefont {{Gong}}}, \bibinfo {author} {\bibfnamefont
  {Z.}~\bibnamefont {{Tu}}}, \bibinfo {author} {\bibfnamefont {H.}~\bibnamefont
  {{Lei}}}, \bibinfo {author} {\bibfnamefont {H.}~\bibnamefont {{Tan}}},
  \bibinfo {author} {\bibfnamefont {S.}~\bibnamefont {{Zhou}}}, \bibinfo
  {author} {\bibfnamefont {C.}~\bibnamefont {{Shen}}}, \bibinfo {author}
  {\bibfnamefont {X.}~\bibnamefont {{Dong}}}, \bibinfo {author} {\bibfnamefont
  {B.}~\bibnamefont {{Yan}}}, \bibinfo {author} {\bibfnamefont
  {Z.}~\bibnamefont {{Wang}}},\ and\ \bibinfo {author} {\bibfnamefont {H.-J.}\
  \bibnamefont {{Gao}}},\ }\bibfield  {title} {\bibinfo {title} {{Roton pair
  density wave in a strong-coupling kagome superconductor}},\ }\href
  {https://doi.org/10.1038/s41586-021-03983-5} {\bibfield  {journal} {\bibinfo
  {journal} {Nature}\ }\textbf {\bibinfo {volume} {599}},\ \bibinfo {pages}
  {222} (\bibinfo {year} {2021})}\BibitemShut {NoStop}%
\bibitem [{\citenamefont {Li}\ \emph {et~al.}(2021)\citenamefont {Li},
  \citenamefont {Zhang}, \citenamefont {Yilmaz}, \citenamefont {Pai},
  \citenamefont {Marvinney}, \citenamefont {Said}, \citenamefont {Yin},
  \citenamefont {Gong}, \citenamefont {Tu}, \citenamefont {Vescovo},
  \citenamefont {Nelson}, \citenamefont {Moore}, \citenamefont {Murakami},
  \citenamefont {Lei}, \citenamefont {Lee}, \citenamefont {Lawrie},\ and\
  \citenamefont {Miao}}]{li21prx}%
  \BibitemOpen
  \bibfield  {author} {\bibinfo {author} {\bibfnamefont {H.}~\bibnamefont
  {Li}}, \bibinfo {author} {\bibfnamefont {T.~T.}\ \bibnamefont {Zhang}},
  \bibinfo {author} {\bibfnamefont {T.}~\bibnamefont {Yilmaz}}, \bibinfo
  {author} {\bibfnamefont {Y.~Y.}\ \bibnamefont {Pai}}, \bibinfo {author}
  {\bibfnamefont {C.~E.}\ \bibnamefont {Marvinney}}, \bibinfo {author}
  {\bibfnamefont {A.}~\bibnamefont {Said}}, \bibinfo {author} {\bibfnamefont
  {Q.~W.}\ \bibnamefont {Yin}}, \bibinfo {author} {\bibfnamefont {C.~S.}\
  \bibnamefont {Gong}}, \bibinfo {author} {\bibfnamefont {Z.~J.}\ \bibnamefont
  {Tu}}, \bibinfo {author} {\bibfnamefont {E.}~\bibnamefont {Vescovo}},
  \bibinfo {author} {\bibfnamefont {C.~S.}\ \bibnamefont {Nelson}}, \bibinfo
  {author} {\bibfnamefont {R.~G.}\ \bibnamefont {Moore}}, \bibinfo {author}
  {\bibfnamefont {S.}~\bibnamefont {Murakami}}, \bibinfo {author}
  {\bibfnamefont {H.~C.}\ \bibnamefont {Lei}}, \bibinfo {author} {\bibfnamefont
  {H.~N.}\ \bibnamefont {Lee}}, \bibinfo {author} {\bibfnamefont {B.~J.}\
  \bibnamefont {Lawrie}},\ and\ \bibinfo {author} {\bibfnamefont
  {H.}~\bibnamefont {Miao}},\ }\bibfield  {title} {\bibinfo {title}
  {{Observation of Unconventional Charge Density Wave without Acoustic Phonon
  Anomaly in Kagome Superconductors ${A\mathrm{V}}_{3}{\mathrm{Sb}}_{5}$
  ($A=\mathrm{Rb}$, Cs)}},\ }\href {https://doi.org/10.1103/PhysRevX.11.031050}
  {\bibfield  {journal} {\bibinfo  {journal} {Phys. Rev. X}\ }\textbf {\bibinfo
  {volume} {11}},\ \bibinfo {pages} {031050} (\bibinfo {year}
  {2021})}\BibitemShut {NoStop}%
\bibitem [{\citenamefont {{Wang}}\ \emph {et~al.}(2021)\citenamefont {{Wang}},
  \citenamefont {{Ma}}, \citenamefont {{Zhang}}, \citenamefont {{Yang}},
  \citenamefont {{Zhao}}, \citenamefont {{Ou}}, \citenamefont {{Zhu}},
  \citenamefont {{Ni}}, \citenamefont {{Lu}}, \citenamefont {{Chen}},
  \citenamefont {{Jiang}}, \citenamefont {{Yu}}, \citenamefont {{Zhang}},
  \citenamefont {{Dong}}, \citenamefont {{Hu}}, \citenamefont {{Gao}},\ and\
  \citenamefont {{Zhao}}}]{wang21ax}%
  \BibitemOpen
  \bibfield  {author} {\bibinfo {author} {\bibfnamefont {Z.}~\bibnamefont
  {{Wang}}}, \bibinfo {author} {\bibfnamefont {S.}~\bibnamefont {{Ma}}},
  \bibinfo {author} {\bibfnamefont {Y.}~\bibnamefont {{Zhang}}}, \bibinfo
  {author} {\bibfnamefont {H.}~\bibnamefont {{Yang}}}, \bibinfo {author}
  {\bibfnamefont {Z.}~\bibnamefont {{Zhao}}}, \bibinfo {author} {\bibfnamefont
  {Y.}~\bibnamefont {{Ou}}}, \bibinfo {author} {\bibfnamefont {Y.}~\bibnamefont
  {{Zhu}}}, \bibinfo {author} {\bibfnamefont {S.}~\bibnamefont {{Ni}}},
  \bibinfo {author} {\bibfnamefont {Z.}~\bibnamefont {{Lu}}}, \bibinfo {author}
  {\bibfnamefont {H.}~\bibnamefont {{Chen}}}, \bibinfo {author} {\bibfnamefont
  {K.}~\bibnamefont {{Jiang}}}, \bibinfo {author} {\bibfnamefont
  {L.}~\bibnamefont {{Yu}}}, \bibinfo {author} {\bibfnamefont {Y.}~\bibnamefont
  {{Zhang}}}, \bibinfo {author} {\bibfnamefont {X.}~\bibnamefont {{Dong}}},
  \bibinfo {author} {\bibfnamefont {J.}~\bibnamefont {{Hu}}}, \bibinfo {author}
  {\bibfnamefont {H.-J.}\ \bibnamefont {{Gao}}},\ and\ \bibinfo {author}
  {\bibfnamefont {Z.}~\bibnamefont {{Zhao}}},\ }\bibfield  {title} {\bibinfo
  {title} {{Distinctive momentum dependent charge-density-wave gap observed in
  CsV$_3$Sb$_5$ superconductor with topological Kagome lattice}},\ }\href@noop
  {} {\bibfield  {journal} {\bibinfo  {journal} {arXiv e-prints}\ ,\ \bibinfo
  {eid} {arXiv:2104.05556}} (\bibinfo {year} {2021})},\ \Eprint
  {https://arxiv.org/abs/2104.05556} {arXiv:2104.05556 [cond-mat.supr-con]}
  \BibitemShut {NoStop}%
\bibitem [{\citenamefont {Nakayama}\ \emph {et~al.}(2021)\citenamefont
  {Nakayama}, \citenamefont {Li}, \citenamefont {Kato}, \citenamefont {Liu},
  \citenamefont {Wang}, \citenamefont {Takahashi}, \citenamefont {Yao},\ and\
  \citenamefont {Sato}}]{nakayama21prb}%
  \BibitemOpen
  \bibfield  {author} {\bibinfo {author} {\bibfnamefont {K.}~\bibnamefont
  {Nakayama}}, \bibinfo {author} {\bibfnamefont {Y.}~\bibnamefont {Li}},
  \bibinfo {author} {\bibfnamefont {T.}~\bibnamefont {Kato}}, \bibinfo {author}
  {\bibfnamefont {M.}~\bibnamefont {Liu}}, \bibinfo {author} {\bibfnamefont
  {Z.}~\bibnamefont {Wang}}, \bibinfo {author} {\bibfnamefont {T.}~\bibnamefont
  {Takahashi}}, \bibinfo {author} {\bibfnamefont {Y.}~\bibnamefont {Yao}},\
  and\ \bibinfo {author} {\bibfnamefont {T.}~\bibnamefont {Sato}},\ }\bibfield
  {title} {\bibinfo {title} {{Multiple energy scales and anisotropic energy gap
  in the charge-density-wave phase of the kagome superconductor
  ${\mathrm{CsV}}_{3}{\mathrm{Sb}}_{5}$}},\ }\href
  {https://doi.org/10.1103/PhysRevB.104.L161112} {\bibfield  {journal}
  {\bibinfo  {journal} {Phys. Rev. B}\ }\textbf {\bibinfo {volume} {104}},\
  \bibinfo {pages} {L161112} (\bibinfo {year} {2021})}\BibitemShut {NoStop}%
\bibitem [{\citenamefont {{Li}}\ \emph {et~al.}(2021)\citenamefont {{Li}},
  \citenamefont {{Zhao}}, \citenamefont {{Ortiz}}, \citenamefont {{Park}},
  \citenamefont {{Ye}}, \citenamefont {{Balents}}, \citenamefont {{Wang}},
  \citenamefont {{Wilson}},\ and\ \citenamefont {{Zeljkovic}}}]{li22np}%
  \BibitemOpen
  \bibfield  {author} {\bibinfo {author} {\bibfnamefont {H.}~\bibnamefont
  {{Li}}}, \bibinfo {author} {\bibfnamefont {H.}~\bibnamefont {{Zhao}}},
  \bibinfo {author} {\bibfnamefont {B.~R.}\ \bibnamefont {{Ortiz}}}, \bibinfo
  {author} {\bibfnamefont {T.}~\bibnamefont {{Park}}}, \bibinfo {author}
  {\bibfnamefont {M.}~\bibnamefont {{Ye}}}, \bibinfo {author} {\bibfnamefont
  {L.}~\bibnamefont {{Balents}}}, \bibinfo {author} {\bibfnamefont
  {Z.}~\bibnamefont {{Wang}}}, \bibinfo {author} {\bibfnamefont {S.~D.}\
  \bibnamefont {{Wilson}}},\ and\ \bibinfo {author} {\bibfnamefont
  {I.}~\bibnamefont {{Zeljkovic}}},\ }\bibfield  {title} {\bibinfo {title}
  {{Rotation symmetry breaking in the normal state of a kagome superconductor
  KV$_3$Sb$_5$}},\ }\href {https://doi.org/10.1038/s41567-021-01479-7}
  {\bibfield  {journal} {\bibinfo  {journal} {Nat. Phys.}\ }\textbf {\bibinfo
  {volume} {18}},\ \bibinfo {pages} {265} (\bibinfo {year} {2021})}\BibitemShut
  {NoStop}%
\bibitem [{\citenamefont {Shumiya}\ \emph {et~al.}(2021)\citenamefont
  {Shumiya}, \citenamefont {Hossain}, \citenamefont {Yin}, \citenamefont
  {Jiang}, \citenamefont {Ortiz}, \citenamefont {Liu}, \citenamefont {Shi},
  \citenamefont {Yin}, \citenamefont {Lei}, \citenamefont {Zhang},
  \citenamefont {Chang}, \citenamefont {Zhang}, \citenamefont {Cochran},
  \citenamefont {Multer}, \citenamefont {Litskevich}, \citenamefont {Cheng},
  \citenamefont {Yang}, \citenamefont {Guguchia}, \citenamefont {Wilson},\ and\
  \citenamefont {Hasan}}]{shumiya21prb}%
  \BibitemOpen
  \bibfield  {author} {\bibinfo {author} {\bibfnamefont {N.}~\bibnamefont
  {Shumiya}}, \bibinfo {author} {\bibfnamefont {M.~S.}\ \bibnamefont
  {Hossain}}, \bibinfo {author} {\bibfnamefont {J.-X.}\ \bibnamefont {Yin}},
  \bibinfo {author} {\bibfnamefont {Y.-X.}\ \bibnamefont {Jiang}}, \bibinfo
  {author} {\bibfnamefont {B.~R.}\ \bibnamefont {Ortiz}}, \bibinfo {author}
  {\bibfnamefont {H.}~\bibnamefont {Liu}}, \bibinfo {author} {\bibfnamefont
  {Y.}~\bibnamefont {Shi}}, \bibinfo {author} {\bibfnamefont {Q.}~\bibnamefont
  {Yin}}, \bibinfo {author} {\bibfnamefont {H.}~\bibnamefont {Lei}}, \bibinfo
  {author} {\bibfnamefont {S.~S.}\ \bibnamefont {Zhang}}, \bibinfo {author}
  {\bibfnamefont {G.}~\bibnamefont {Chang}}, \bibinfo {author} {\bibfnamefont
  {Q.}~\bibnamefont {Zhang}}, \bibinfo {author} {\bibfnamefont {T.~A.}\
  \bibnamefont {Cochran}}, \bibinfo {author} {\bibfnamefont {D.}~\bibnamefont
  {Multer}}, \bibinfo {author} {\bibfnamefont {M.}~\bibnamefont {Litskevich}},
  \bibinfo {author} {\bibfnamefont {Z.-J.}\ \bibnamefont {Cheng}}, \bibinfo
  {author} {\bibfnamefont {X.~P.}\ \bibnamefont {Yang}}, \bibinfo {author}
  {\bibfnamefont {Z.}~\bibnamefont {Guguchia}}, \bibinfo {author}
  {\bibfnamefont {S.~D.}\ \bibnamefont {Wilson}},\ and\ \bibinfo {author}
  {\bibfnamefont {M.~Z.}\ \bibnamefont {Hasan}},\ }\bibfield  {title} {\bibinfo
  {title} {{Intrinsic nature of chiral charge order in the kagome
  superconductor $\mathrm{Rb}{\mathrm{V}}_{3}{\mathrm{Sb}}_{5}$}},\ }\href
  {https://doi.org/10.1103/PhysRevB.104.035131} {\bibfield  {journal} {\bibinfo
   {journal} {Phys. Rev. B}\ }\textbf {\bibinfo {volume} {104}},\ \bibinfo
  {pages} {035131} (\bibinfo {year} {2021})}\BibitemShut {NoStop}%
\bibitem [{\citenamefont {Tsirlin}\ \emph {et~al.}(2022)\citenamefont
  {Tsirlin}, \citenamefont {Fertey}, \citenamefont {Ortiz}, \citenamefont
  {Klis}, \citenamefont {Merkl}, \citenamefont {Dressel}, \citenamefont
  {Wilson},\ and\ \citenamefont {Uykur}}]{tsirlin21sp}%
  \BibitemOpen
  \bibfield  {author} {\bibinfo {author} {\bibfnamefont {A.~A.}\ \bibnamefont
  {Tsirlin}}, \bibinfo {author} {\bibfnamefont {P.}~\bibnamefont {Fertey}},
  \bibinfo {author} {\bibfnamefont {B.~R.}\ \bibnamefont {Ortiz}}, \bibinfo
  {author} {\bibfnamefont {B.}~\bibnamefont {Klis}}, \bibinfo {author}
  {\bibfnamefont {V.}~\bibnamefont {Merkl}}, \bibinfo {author} {\bibfnamefont
  {M.}~\bibnamefont {Dressel}}, \bibinfo {author} {\bibfnamefont {S.~D.}\
  \bibnamefont {Wilson}},\ and\ \bibinfo {author} {\bibfnamefont
  {E.}~\bibnamefont {Uykur}},\ }\bibfield  {title} {\bibinfo {title} {{Role of
  Sb in the superconducting kagome metal CsV$_3$Sb$_5$ revealed by its
  anisotropic compression}},\ }\href
  {https://doi.org/10.21468/SciPostPhys.12.2.049} {\bibfield  {journal}
  {\bibinfo  {journal} {SciPost Phys.}\ }\textbf {\bibinfo {volume} {12}},\
  \bibinfo {pages} {49} (\bibinfo {year} {2022})}\BibitemShut {NoStop}%
\bibitem [{\citenamefont {{Kang}}\ \emph {et~al.}(2022)\citenamefont {{Kang}},
  \citenamefont {{Fang}}, \citenamefont {{Kim}}, \citenamefont {{Ortiz}},
  \citenamefont {{Ryu}}, \citenamefont {{Kim}}, \citenamefont {{Yoo}},
  \citenamefont {{Sangiovanni}}, \citenamefont {{Di Sante}}, \citenamefont
  {{Park}}, \citenamefont {{Jozwiak}}, \citenamefont {{Bostwick}},
  \citenamefont {{Rotenberg}}, \citenamefont {{Kaxiras}}, \citenamefont
  {{Wilson}}, \citenamefont {{Park}},\ and\ \citenamefont
  {{Comin}}}]{kang22np}%
  \BibitemOpen
  \bibfield  {author} {\bibinfo {author} {\bibfnamefont {M.}~\bibnamefont
  {{Kang}}}, \bibinfo {author} {\bibfnamefont {S.}~\bibnamefont {{Fang}}},
  \bibinfo {author} {\bibfnamefont {J.-K.}\ \bibnamefont {{Kim}}}, \bibinfo
  {author} {\bibfnamefont {B.~R.}\ \bibnamefont {{Ortiz}}}, \bibinfo {author}
  {\bibfnamefont {S.~H.}\ \bibnamefont {{Ryu}}}, \bibinfo {author}
  {\bibfnamefont {J.}~\bibnamefont {{Kim}}}, \bibinfo {author} {\bibfnamefont
  {J.}~\bibnamefont {{Yoo}}}, \bibinfo {author} {\bibfnamefont
  {G.}~\bibnamefont {{Sangiovanni}}}, \bibinfo {author} {\bibfnamefont
  {D.}~\bibnamefont {{Di Sante}}}, \bibinfo {author} {\bibfnamefont {B.-G.}\
  \bibnamefont {{Park}}}, \bibinfo {author} {\bibfnamefont {C.}~\bibnamefont
  {{Jozwiak}}}, \bibinfo {author} {\bibfnamefont {A.}~\bibnamefont
  {{Bostwick}}}, \bibinfo {author} {\bibfnamefont {E.}~\bibnamefont
  {{Rotenberg}}}, \bibinfo {author} {\bibfnamefont {E.}~\bibnamefont
  {{Kaxiras}}}, \bibinfo {author} {\bibfnamefont {S.~D.}\ \bibnamefont
  {{Wilson}}}, \bibinfo {author} {\bibfnamefont {J.-H.}\ \bibnamefont
  {{Park}}},\ and\ \bibinfo {author} {\bibfnamefont {R.}~\bibnamefont
  {{Comin}}},\ }\bibfield  {title} {\bibinfo {title} {{Twofold van Hove
  singularity and origin of charge order in topological kagome superconductor
  CsV$_{3}$Sb$_{5}$}},\ }\href {https://doi.org/10.1038/s41567-021-01451-5}
  {\bibfield  {journal} {\bibinfo  {journal} {Nat. Phys.}\ }\textbf {\bibinfo
  {volume} {18}},\ \bibinfo {pages} {301} (\bibinfo {year} {2022})}\BibitemShut
  {NoStop}%
\bibitem [{\citenamefont {Wang}\ \emph
  {et~al.}(2021{\natexlab{a}})\citenamefont {Wang}, \citenamefont {Jiang},
  \citenamefont {Yin}, \citenamefont {Li}, \citenamefont {Wang}, \citenamefont
  {Huang}, \citenamefont {Shao}, \citenamefont {Liu}, \citenamefont {Zhu},
  \citenamefont {Shumiya}, \citenamefont {Hossain}, \citenamefont {Liu},
  \citenamefont {Shi}, \citenamefont {Duan}, \citenamefont {Li}, \citenamefont
  {Chang}, \citenamefont {Dai}, \citenamefont {Ye}, \citenamefont {Xu},
  \citenamefont {Wang}, \citenamefont {Zheng}, \citenamefont {Jia},
  \citenamefont {Hasan},\ and\ \citenamefont {Yao}}]{wang21prb}%
  \BibitemOpen
  \bibfield  {author} {\bibinfo {author} {\bibfnamefont {Z.}~\bibnamefont
  {Wang}}, \bibinfo {author} {\bibfnamefont {Y.-X.}\ \bibnamefont {Jiang}},
  \bibinfo {author} {\bibfnamefont {J.-X.}\ \bibnamefont {Yin}}, \bibinfo
  {author} {\bibfnamefont {Y.}~\bibnamefont {Li}}, \bibinfo {author}
  {\bibfnamefont {G.-Y.}\ \bibnamefont {Wang}}, \bibinfo {author}
  {\bibfnamefont {H.-L.}\ \bibnamefont {Huang}}, \bibinfo {author}
  {\bibfnamefont {S.}~\bibnamefont {Shao}}, \bibinfo {author} {\bibfnamefont
  {J.}~\bibnamefont {Liu}}, \bibinfo {author} {\bibfnamefont {P.}~\bibnamefont
  {Zhu}}, \bibinfo {author} {\bibfnamefont {N.}~\bibnamefont {Shumiya}},
  \bibinfo {author} {\bibfnamefont {M.~S.}\ \bibnamefont {Hossain}}, \bibinfo
  {author} {\bibfnamefont {H.}~\bibnamefont {Liu}}, \bibinfo {author}
  {\bibfnamefont {Y.}~\bibnamefont {Shi}}, \bibinfo {author} {\bibfnamefont
  {J.}~\bibnamefont {Duan}}, \bibinfo {author} {\bibfnamefont {X.}~\bibnamefont
  {Li}}, \bibinfo {author} {\bibfnamefont {G.}~\bibnamefont {Chang}}, \bibinfo
  {author} {\bibfnamefont {P.}~\bibnamefont {Dai}}, \bibinfo {author}
  {\bibfnamefont {Z.}~\bibnamefont {Ye}}, \bibinfo {author} {\bibfnamefont
  {G.}~\bibnamefont {Xu}}, \bibinfo {author} {\bibfnamefont {Y.}~\bibnamefont
  {Wang}}, \bibinfo {author} {\bibfnamefont {H.}~\bibnamefont {Zheng}},
  \bibinfo {author} {\bibfnamefont {J.}~\bibnamefont {Jia}}, \bibinfo {author}
  {\bibfnamefont {M.~Z.}\ \bibnamefont {Hasan}},\ and\ \bibinfo {author}
  {\bibfnamefont {Y.}~\bibnamefont {Yao}},\ }\bibfield  {title} {\bibinfo
  {title} {{Electronic nature of chiral charge order in the kagome
  superconductor $\mathrm{Cs}{\mathrm{V}}_{3}{\mathrm{Sb}}_{5}$}},\ }\href
  {https://doi.org/10.1103/PhysRevB.104.075148} {\bibfield  {journal} {\bibinfo
   {journal} {Phys. Rev. B}\ }\textbf {\bibinfo {volume} {104}},\ \bibinfo
  {pages} {075148} (\bibinfo {year} {2021}{\natexlab{a}})}\BibitemShut
  {NoStop}%
\bibitem [{\citenamefont {Cho}\ \emph {et~al.}(2021)\citenamefont {Cho},
  \citenamefont {Ma}, \citenamefont {Xia}, \citenamefont {Yang}, \citenamefont
  {Liu}, \citenamefont {Huang}, \citenamefont {Jiang}, \citenamefont {Lu},
  \citenamefont {Liu}, \citenamefont {Liu}, \citenamefont {Li}, \citenamefont
  {Wang}, \citenamefont {Liu}, \citenamefont {Jia}, \citenamefont {Guo},
  \citenamefont {Liu},\ and\ \citenamefont {Shen}}]{cho21prl}%
  \BibitemOpen
  \bibfield  {author} {\bibinfo {author} {\bibfnamefont {S.}~\bibnamefont
  {Cho}}, \bibinfo {author} {\bibfnamefont {H.}~\bibnamefont {Ma}}, \bibinfo
  {author} {\bibfnamefont {W.}~\bibnamefont {Xia}}, \bibinfo {author}
  {\bibfnamefont {Y.}~\bibnamefont {Yang}}, \bibinfo {author} {\bibfnamefont
  {Z.}~\bibnamefont {Liu}}, \bibinfo {author} {\bibfnamefont {Z.}~\bibnamefont
  {Huang}}, \bibinfo {author} {\bibfnamefont {Z.}~\bibnamefont {Jiang}},
  \bibinfo {author} {\bibfnamefont {X.}~\bibnamefont {Lu}}, \bibinfo {author}
  {\bibfnamefont {J.}~\bibnamefont {Liu}}, \bibinfo {author} {\bibfnamefont
  {Z.}~\bibnamefont {Liu}}, \bibinfo {author} {\bibfnamefont {J.}~\bibnamefont
  {Li}}, \bibinfo {author} {\bibfnamefont {J.}~\bibnamefont {Wang}}, \bibinfo
  {author} {\bibfnamefont {Y.}~\bibnamefont {Liu}}, \bibinfo {author}
  {\bibfnamefont {J.}~\bibnamefont {Jia}}, \bibinfo {author} {\bibfnamefont
  {Y.}~\bibnamefont {Guo}}, \bibinfo {author} {\bibfnamefont {J.}~\bibnamefont
  {Liu}},\ and\ \bibinfo {author} {\bibfnamefont {D.}~\bibnamefont {Shen}},\
  }\bibfield  {title} {\bibinfo {title} {{Emergence of New van Hove
  Singularities in the Charge Density Wave State of a Topological Kagome Metal
  ${\mathrm{RbV}}_{3}{\mathrm{Sb}}_{5}$}},\ }\href
  {https://doi.org/10.1103/PhysRevLett.127.236401} {\bibfield  {journal}
  {\bibinfo  {journal} {Phys. Rev. Lett.}\ }\textbf {\bibinfo {volume} {127}},\
  \bibinfo {pages} {236401} (\bibinfo {year} {2021})}\BibitemShut {NoStop}%
\bibitem [{\citenamefont {{Song}}\ \emph {et~al.}(2021)\citenamefont {{Song}},
  \citenamefont {{Kong}}, \citenamefont {{Xia}}, \citenamefont {{Yin}},
  \citenamefont {{Tu}}, \citenamefont {{Zhao}}, \citenamefont {{Dai}},
  \citenamefont {{Meng}}, \citenamefont {{Tao}}, \citenamefont {{Tu}},
  \citenamefont {{Gong}}, \citenamefont {{Lei}}, \citenamefont {{Guo}},
  \citenamefont {{Yang}},\ and\ \citenamefont {{Li}}}]{song21ax}%
  \BibitemOpen
  \bibfield  {author} {\bibinfo {author} {\bibfnamefont {B.~Q.}\ \bibnamefont
  {{Song}}}, \bibinfo {author} {\bibfnamefont {X.~M.}\ \bibnamefont {{Kong}}},
  \bibinfo {author} {\bibfnamefont {W.}~\bibnamefont {{Xia}}}, \bibinfo
  {author} {\bibfnamefont {Q.~W.}\ \bibnamefont {{Yin}}}, \bibinfo {author}
  {\bibfnamefont {C.~P.}\ \bibnamefont {{Tu}}}, \bibinfo {author}
  {\bibfnamefont {C.~C.}\ \bibnamefont {{Zhao}}}, \bibinfo {author}
  {\bibfnamefont {D.~Z.}\ \bibnamefont {{Dai}}}, \bibinfo {author}
  {\bibfnamefont {K.}~\bibnamefont {{Meng}}}, \bibinfo {author} {\bibfnamefont
  {Z.~C.}\ \bibnamefont {{Tao}}}, \bibinfo {author} {\bibfnamefont {Z.~J.}\
  \bibnamefont {{Tu}}}, \bibinfo {author} {\bibfnamefont {C.~S.}\ \bibnamefont
  {{Gong}}}, \bibinfo {author} {\bibfnamefont {H.~C.}\ \bibnamefont {{Lei}}},
  \bibinfo {author} {\bibfnamefont {Y.~F.}\ \bibnamefont {{Guo}}}, \bibinfo
  {author} {\bibfnamefont {X.~F.}\ \bibnamefont {{Yang}}},\ and\ \bibinfo
  {author} {\bibfnamefont {S.~Y.}\ \bibnamefont {{Li}}},\ }\bibfield  {title}
  {\bibinfo {title} {{Competing superconductivity and charge-density wave in
  Kagome metal CsV3Sb5: evidence from their evolutions with sample
  thickness}},\ }\href@noop {} {\bibfield  {journal} {\bibinfo  {journal}
  {arXiv e-prints}\ ,\ \bibinfo {eid} {arXiv:2105.09248}} (\bibinfo {year}
  {2021})},\ \Eprint {https://arxiv.org/abs/2105.09248} {arXiv:2105.09248
  [cond-mat.supr-con]} \BibitemShut {NoStop}%
\bibitem [{\citenamefont {Song}\ \emph {et~al.}(2021)\citenamefont {Song},
  \citenamefont {Ying}, \citenamefont {Chen}, \citenamefont {Han},
  \citenamefont {Wu}, \citenamefont {Schnyder}, \citenamefont {Huang},
  \citenamefont {Guo},\ and\ \citenamefont {Chen}}]{song21prl}%
  \BibitemOpen
  \bibfield  {author} {\bibinfo {author} {\bibfnamefont {Y.}~\bibnamefont
  {Song}}, \bibinfo {author} {\bibfnamefont {T.}~\bibnamefont {Ying}}, \bibinfo
  {author} {\bibfnamefont {X.}~\bibnamefont {Chen}}, \bibinfo {author}
  {\bibfnamefont {X.}~\bibnamefont {Han}}, \bibinfo {author} {\bibfnamefont
  {X.}~\bibnamefont {Wu}}, \bibinfo {author} {\bibfnamefont {A.~P.}\
  \bibnamefont {Schnyder}}, \bibinfo {author} {\bibfnamefont {Y.}~\bibnamefont
  {Huang}}, \bibinfo {author} {\bibfnamefont {J.-g.}\ \bibnamefont {Guo}},\
  and\ \bibinfo {author} {\bibfnamefont {X.}~\bibnamefont {Chen}},\ }\bibfield
  {title} {\bibinfo {title} {{Competition of Superconductivity and Charge
  Density Wave in Selective Oxidized ${\mathrm{CsV}}_{3}{\mathrm{Sb}}_{5}$ Thin
  Flakes}},\ }\href {https://doi.org/10.1103/PhysRevLett.127.237001} {\bibfield
   {journal} {\bibinfo  {journal} {Phys. Rev. Lett.}\ }\textbf {\bibinfo
  {volume} {127}},\ \bibinfo {pages} {237001} (\bibinfo {year}
  {2021})}\BibitemShut {NoStop}%
\bibitem [{\citenamefont {Yu}\ \emph {et~al.}(2021{\natexlab{b}})\citenamefont
  {Yu}, \citenamefont {Ma}, \citenamefont {Zhuo}, \citenamefont {Liu},
  \citenamefont {Wen}, \citenamefont {Lei}, \citenamefont {Ying},\ and\
  \citenamefont {Chen}}]{yu21nc}%
  \BibitemOpen
  \bibfield  {author} {\bibinfo {author} {\bibfnamefont {F.~H.}\ \bibnamefont
  {Yu}}, \bibinfo {author} {\bibfnamefont {D.~H.}\ \bibnamefont {Ma}}, \bibinfo
  {author} {\bibfnamefont {W.~Z.}\ \bibnamefont {Zhuo}}, \bibinfo {author}
  {\bibfnamefont {S.~Q.}\ \bibnamefont {Liu}}, \bibinfo {author} {\bibfnamefont
  {X.~K.}\ \bibnamefont {Wen}}, \bibinfo {author} {\bibfnamefont
  {B.}~\bibnamefont {Lei}}, \bibinfo {author} {\bibfnamefont {J.~J.}\
  \bibnamefont {Ying}},\ and\ \bibinfo {author} {\bibfnamefont {X.~H.}\
  \bibnamefont {Chen}},\ }\bibfield  {title} {\bibinfo {title} {Unusual
  competition of superconductivity and charge-density-wave state in a
  compressed topological kagome metal},\ }\href
  {https://doi.org/10.1038/s41467-021-23928-w} {\bibfield  {journal} {\bibinfo
  {journal} {Nat. Commun.}\ }\textbf {\bibinfo {volume} {12}},\ \bibinfo
  {pages} {3645} (\bibinfo {year} {2021}{\natexlab{b}})}\BibitemShut {NoStop}%
\bibitem [{\citenamefont {{Hu}}\ \emph {et~al.}(2022)\citenamefont {{Hu}},
  \citenamefont {{Wu}}, \citenamefont {{Ortiz}}, \citenamefont {{Ju}},
  \citenamefont {{Han}}, \citenamefont {{Ma}}, \citenamefont {{Plumb}},
  \citenamefont {{Radovic}}, \citenamefont {{Thomale}}, \citenamefont
  {{Wilson}}, \citenamefont {{Schnyder}},\ and\ \citenamefont
  {{Shi}}}]{hu22nc}%
  \BibitemOpen
  \bibfield  {author} {\bibinfo {author} {\bibfnamefont {Y.}~\bibnamefont
  {{Hu}}}, \bibinfo {author} {\bibfnamefont {X.}~\bibnamefont {{Wu}}}, \bibinfo
  {author} {\bibfnamefont {B.~R.}\ \bibnamefont {{Ortiz}}}, \bibinfo {author}
  {\bibfnamefont {S.}~\bibnamefont {{Ju}}}, \bibinfo {author} {\bibfnamefont
  {X.}~\bibnamefont {{Han}}}, \bibinfo {author} {\bibfnamefont
  {J.}~\bibnamefont {{Ma}}}, \bibinfo {author} {\bibfnamefont {N.~C.}\
  \bibnamefont {{Plumb}}}, \bibinfo {author} {\bibfnamefont {M.}~\bibnamefont
  {{Radovic}}}, \bibinfo {author} {\bibfnamefont {R.}~\bibnamefont
  {{Thomale}}}, \bibinfo {author} {\bibfnamefont {S.~D.}\ \bibnamefont
  {{Wilson}}}, \bibinfo {author} {\bibfnamefont {A.~P.}\ \bibnamefont
  {{Schnyder}}},\ and\ \bibinfo {author} {\bibfnamefont {M.}~\bibnamefont
  {{Shi}}},\ }\bibfield  {title} {\bibinfo {title} {{Rich nature of Van Hove
  singularities in Kagome superconductor CsV$_{3}$Sb$_{5}$}},\ }\href
  {https://doi.org/10.1038/s41467-022-29828-x} {\bibfield  {journal} {\bibinfo
  {journal} {Nat. Commun.}\ }\textbf {\bibinfo {volume} {13}},\ \bibinfo
  {pages} {2220} (\bibinfo {year} {2022})}\BibitemShut {NoStop}%
\bibitem [{\citenamefont {{Mielke}}\ \emph {et~al.}(2022)\citenamefont
  {{Mielke}}, \citenamefont {{Das}}, \citenamefont {{Yin}}, \citenamefont
  {{Liu}}, \citenamefont {{Gupta}}, \citenamefont {{Jiang}}, \citenamefont
  {{Medarde}}, \citenamefont {{Wu}}, \citenamefont {{Lei}}, \citenamefont
  {{Chang}}, \citenamefont {{Dai}}, \citenamefont {{Si}}, \citenamefont
  {{Miao}}, \citenamefont {{Thomale}}, \citenamefont {{Neupert}}, \citenamefont
  {{Shi}}, \citenamefont {{Khasanov}}, \citenamefont {{Hasan}}, \citenamefont
  {{Luetkens}},\ and\ \citenamefont {{Guguchia}}}]{mielke22n}%
  \BibitemOpen
  \bibfield  {author} {\bibinfo {author} {\bibfnamefont {C.}~\bibnamefont
  {{Mielke}}}, \bibinfo {author} {\bibfnamefont {D.}~\bibnamefont {{Das}}},
  \bibinfo {author} {\bibfnamefont {J.~X.}\ \bibnamefont {{Yin}}}, \bibinfo
  {author} {\bibfnamefont {H.}~\bibnamefont {{Liu}}}, \bibinfo {author}
  {\bibfnamefont {R.}~\bibnamefont {{Gupta}}}, \bibinfo {author} {\bibfnamefont
  {Y.~X.}\ \bibnamefont {{Jiang}}}, \bibinfo {author} {\bibfnamefont
  {M.}~\bibnamefont {{Medarde}}}, \bibinfo {author} {\bibfnamefont
  {X.}~\bibnamefont {{Wu}}}, \bibinfo {author} {\bibfnamefont {H.~C.}\
  \bibnamefont {{Lei}}}, \bibinfo {author} {\bibfnamefont {J.}~\bibnamefont
  {{Chang}}}, \bibinfo {author} {\bibfnamefont {P.}~\bibnamefont {{Dai}}},
  \bibinfo {author} {\bibfnamefont {Q.}~\bibnamefont {{Si}}}, \bibinfo {author}
  {\bibfnamefont {H.}~\bibnamefont {{Miao}}}, \bibinfo {author} {\bibfnamefont
  {R.}~\bibnamefont {{Thomale}}}, \bibinfo {author} {\bibfnamefont
  {T.}~\bibnamefont {{Neupert}}}, \bibinfo {author} {\bibfnamefont
  {Y.}~\bibnamefont {{Shi}}}, \bibinfo {author} {\bibfnamefont
  {R.}~\bibnamefont {{Khasanov}}}, \bibinfo {author} {\bibfnamefont {M.~Z.}\
  \bibnamefont {{Hasan}}}, \bibinfo {author} {\bibfnamefont {H.}~\bibnamefont
  {{Luetkens}}},\ and\ \bibinfo {author} {\bibfnamefont {Z.}~\bibnamefont
  {{Guguchia}}},\ }\bibfield  {title} {\bibinfo {title} {{Time-reversal
  symmetry-breaking charge order in a kagome superconductor}},\ }\href
  {https://doi.org/10.1038/s41586-021-04327-z} {\bibfield  {journal} {\bibinfo
  {journal} {Nature}\ }\textbf {\bibinfo {volume} {602}},\ \bibinfo {pages}
  {245} (\bibinfo {year} {2022})}\BibitemShut {NoStop}%
\bibitem [{\citenamefont {Wang}\ \emph
  {et~al.}(2021{\natexlab{b}})\citenamefont {Wang}, \citenamefont {Chen},
  \citenamefont {Yin}, \citenamefont {Ma}, \citenamefont {Pan}, \citenamefont
  {Yang}, \citenamefont {Ji}, \citenamefont {Wu}, \citenamefont {Shan},
  \citenamefont {Xu}, \citenamefont {Tu}, \citenamefont {Gong}, \citenamefont
  {Liu}, \citenamefont {Li}, \citenamefont {Uwatoko}, \citenamefont {Dong},
  \citenamefont {Lei}, \citenamefont {Sun},\ and\ \citenamefont
  {Cheng}}]{wang21prr}%
  \BibitemOpen
  \bibfield  {author} {\bibinfo {author} {\bibfnamefont {N.~N.}\ \bibnamefont
  {Wang}}, \bibinfo {author} {\bibfnamefont {K.~Y.}\ \bibnamefont {Chen}},
  \bibinfo {author} {\bibfnamefont {Q.~W.}\ \bibnamefont {Yin}}, \bibinfo
  {author} {\bibfnamefont {Y.~N.~N.}\ \bibnamefont {Ma}}, \bibinfo {author}
  {\bibfnamefont {B.~Y.}\ \bibnamefont {Pan}}, \bibinfo {author} {\bibfnamefont
  {X.}~\bibnamefont {Yang}}, \bibinfo {author} {\bibfnamefont {X.~Y.}\
  \bibnamefont {Ji}}, \bibinfo {author} {\bibfnamefont {S.~L.}\ \bibnamefont
  {Wu}}, \bibinfo {author} {\bibfnamefont {P.~F.}\ \bibnamefont {Shan}},
  \bibinfo {author} {\bibfnamefont {S.~X.}\ \bibnamefont {Xu}}, \bibinfo
  {author} {\bibfnamefont {Z.~J.}\ \bibnamefont {Tu}}, \bibinfo {author}
  {\bibfnamefont {C.~S.}\ \bibnamefont {Gong}}, \bibinfo {author}
  {\bibfnamefont {G.~T.}\ \bibnamefont {Liu}}, \bibinfo {author} {\bibfnamefont
  {G.}~\bibnamefont {Li}}, \bibinfo {author} {\bibfnamefont {Y.}~\bibnamefont
  {Uwatoko}}, \bibinfo {author} {\bibfnamefont {X.~L.}\ \bibnamefont {Dong}},
  \bibinfo {author} {\bibfnamefont {H.~C.}\ \bibnamefont {Lei}}, \bibinfo
  {author} {\bibfnamefont {J.~P.}\ \bibnamefont {Sun}},\ and\ \bibinfo {author}
  {\bibfnamefont {J.-G.}\ \bibnamefont {Cheng}},\ }\bibfield  {title} {\bibinfo
  {title} {Competition between charge-density-wave and superconductivity in the
  kagome metal $\mathrm{Rb}{\mathrm{v}}_{3}{\mathrm{sb}}_{5}$},\ }\href
  {https://doi.org/10.1103/PhysRevResearch.3.043018} {\bibfield  {journal}
  {\bibinfo  {journal} {Phys. Rev. Research}\ }\textbf {\bibinfo {volume}
  {3}},\ \bibinfo {pages} {043018} (\bibinfo {year}
  {2021}{\natexlab{b}})}\BibitemShut {NoStop}%
\bibitem [{\citenamefont {{Luo}}\ \emph {et~al.}(2022)\citenamefont {{Luo}},
  \citenamefont {{Gao}}, \citenamefont {{Liu}}, \citenamefont {{Gu}},
  \citenamefont {{Wu}}, \citenamefont {{Yi}}, \citenamefont {{Jia}},
  \citenamefont {{Wu}}, \citenamefont {{Luo}}, \citenamefont {{Xu}},
  \citenamefont {{Zhao}}, \citenamefont {{Wang}}, \citenamefont {{Mao}},
  \citenamefont {{Liu}}, \citenamefont {{Zhu}}, \citenamefont {{Shi}},
  \citenamefont {{Jiang}}, \citenamefont {{Hu}}, \citenamefont {{Xu}},\ and\
  \citenamefont {{Zhou}}}]{luo22nc}%
  \BibitemOpen
  \bibfield  {author} {\bibinfo {author} {\bibfnamefont {H.}~\bibnamefont
  {{Luo}}}, \bibinfo {author} {\bibfnamefont {Q.}~\bibnamefont {{Gao}}},
  \bibinfo {author} {\bibfnamefont {H.}~\bibnamefont {{Liu}}}, \bibinfo
  {author} {\bibfnamefont {Y.}~\bibnamefont {{Gu}}}, \bibinfo {author}
  {\bibfnamefont {D.}~\bibnamefont {{Wu}}}, \bibinfo {author} {\bibfnamefont
  {C.}~\bibnamefont {{Yi}}}, \bibinfo {author} {\bibfnamefont {J.}~\bibnamefont
  {{Jia}}}, \bibinfo {author} {\bibfnamefont {S.}~\bibnamefont {{Wu}}},
  \bibinfo {author} {\bibfnamefont {X.}~\bibnamefont {{Luo}}}, \bibinfo
  {author} {\bibfnamefont {Y.}~\bibnamefont {{Xu}}}, \bibinfo {author}
  {\bibfnamefont {L.}~\bibnamefont {{Zhao}}}, \bibinfo {author} {\bibfnamefont
  {Q.}~\bibnamefont {{Wang}}}, \bibinfo {author} {\bibfnamefont
  {H.}~\bibnamefont {{Mao}}}, \bibinfo {author} {\bibfnamefont
  {G.}~\bibnamefont {{Liu}}}, \bibinfo {author} {\bibfnamefont
  {Z.}~\bibnamefont {{Zhu}}}, \bibinfo {author} {\bibfnamefont
  {Y.}~\bibnamefont {{Shi}}}, \bibinfo {author} {\bibfnamefont
  {K.}~\bibnamefont {{Jiang}}}, \bibinfo {author} {\bibfnamefont
  {J.}~\bibnamefont {{Hu}}}, \bibinfo {author} {\bibfnamefont {Z.}~\bibnamefont
  {{Xu}}},\ and\ \bibinfo {author} {\bibfnamefont {X.~J.}\ \bibnamefont
  {{Zhou}}},\ }\bibfield  {title} {\bibinfo {title} {{Electronic nature of
  charge density wave and electron-phonon coupling in kagome superconductor
  KV$_{3}$Sb$_{5}$}},\ }\href {https://doi.org/10.1038/s41467-021-27946-6}
  {\bibfield  {journal} {\bibinfo  {journal} {Nat. Commun.}\ }\textbf {\bibinfo
  {volume} {13}},\ \bibinfo {pages} {273} (\bibinfo {year} {2022})}\BibitemShut
  {NoStop}%
\bibitem [{\citenamefont {Qian}\ \emph {et~al.}(2021)\citenamefont {Qian},
  \citenamefont {Christensen}, \citenamefont {Hu}, \citenamefont {Saha},
  \citenamefont {Andersen}, \citenamefont {Fernandes}, \citenamefont {Birol},\
  and\ \citenamefont {Ni}}]{qian21prb}%
  \BibitemOpen
  \bibfield  {author} {\bibinfo {author} {\bibfnamefont {T.}~\bibnamefont
  {Qian}}, \bibinfo {author} {\bibfnamefont {M.~H.}\ \bibnamefont
  {Christensen}}, \bibinfo {author} {\bibfnamefont {C.}~\bibnamefont {Hu}},
  \bibinfo {author} {\bibfnamefont {A.}~\bibnamefont {Saha}}, \bibinfo {author}
  {\bibfnamefont {B.~M.}\ \bibnamefont {Andersen}}, \bibinfo {author}
  {\bibfnamefont {R.~M.}\ \bibnamefont {Fernandes}}, \bibinfo {author}
  {\bibfnamefont {T.}~\bibnamefont {Birol}},\ and\ \bibinfo {author}
  {\bibfnamefont {N.}~\bibnamefont {Ni}},\ }\bibfield  {title} {\bibinfo
  {title} {Revealing the competition between charge density wave and
  superconductivity in ${{\mathrm{CsV}}_{3}\mathrm{Sb}}_{5}$ through uniaxial
  strain},\ }\href {https://doi.org/10.1103/PhysRevB.104.144506} {\bibfield
  {journal} {\bibinfo  {journal} {Phys. Rev. B}\ }\textbf {\bibinfo {volume}
  {104}},\ \bibinfo {pages} {144506} (\bibinfo {year} {2021})}\BibitemShut
  {NoStop}%
\bibitem [{\citenamefont {Van~Hove}(1953)}]{vanhove53pr}%
  \BibitemOpen
  \bibfield  {author} {\bibinfo {author} {\bibfnamefont {L.}~\bibnamefont
  {Van~Hove}},\ }\bibfield  {title} {\bibinfo {title} {The occurrence of
  singularities in the elastic frequency distribution of a crystal},\ }\href
  {https://doi.org/10.1103/PhysRev.89.1189} {\bibfield  {journal} {\bibinfo
  {journal} {Phys. Rev.}\ }\textbf {\bibinfo {volume} {89}},\ \bibinfo {pages}
  {1189} (\bibinfo {year} {1953})}\BibitemShut {NoStop}%
\bibitem [{\citenamefont {Tan}\ \emph {et~al.}(2021)\citenamefont {Tan},
  \citenamefont {Liu}, \citenamefont {Wang},\ and\ \citenamefont
  {Yan}}]{tan21prl}%
  \BibitemOpen
  \bibfield  {author} {\bibinfo {author} {\bibfnamefont {H.}~\bibnamefont
  {Tan}}, \bibinfo {author} {\bibfnamefont {Y.}~\bibnamefont {Liu}}, \bibinfo
  {author} {\bibfnamefont {Z.}~\bibnamefont {Wang}},\ and\ \bibinfo {author}
  {\bibfnamefont {B.}~\bibnamefont {Yan}},\ }\bibfield  {title} {\bibinfo
  {title} {Charge density waves and electronic properties of superconducting
  kagome metals},\ }\href {https://doi.org/10.1103/PhysRevLett.127.046401}
  {\bibfield  {journal} {\bibinfo  {journal} {Phys. Rev. Lett.}\ }\textbf
  {\bibinfo {volume} {127}},\ \bibinfo {pages} {046401} (\bibinfo {year}
  {2021})}\BibitemShut {NoStop}%
\bibitem [{\citenamefont {Feng}\ \emph
  {et~al.}(2021{\natexlab{a}})\citenamefont {Feng}, \citenamefont {Jiang},
  \citenamefont {Wang},\ and\ \citenamefont {Hu}}]{feng21sb}%
  \BibitemOpen
  \bibfield  {author} {\bibinfo {author} {\bibfnamefont {X.}~\bibnamefont
  {Feng}}, \bibinfo {author} {\bibfnamefont {K.}~\bibnamefont {Jiang}},
  \bibinfo {author} {\bibfnamefont {Z.}~\bibnamefont {Wang}},\ and\ \bibinfo
  {author} {\bibfnamefont {J.}~\bibnamefont {Hu}},\ }\bibfield  {title}
  {\bibinfo {title} {{Chiral flux phase in the Kagome superconductor
  AV$_3$Sb$_5$}},\ }\href
  {https://doi.org/https://doi.org/10.1016/j.scib.2021.04.043} {\bibfield
  {journal} {\bibinfo  {journal} {Sci. Bull.}\ }\textbf {\bibinfo {volume}
  {66}},\ \bibinfo {pages} {1384} (\bibinfo {year}
  {2021}{\natexlab{a}})}\BibitemShut {NoStop}%
\bibitem [{\citenamefont {Denner}\ \emph {et~al.}(2021)\citenamefont {Denner},
  \citenamefont {Thomale},\ and\ \citenamefont {Neupert}}]{denner21prl}%
  \BibitemOpen
  \bibfield  {author} {\bibinfo {author} {\bibfnamefont {M.~M.}\ \bibnamefont
  {Denner}}, \bibinfo {author} {\bibfnamefont {R.}~\bibnamefont {Thomale}},\
  and\ \bibinfo {author} {\bibfnamefont {T.}~\bibnamefont {Neupert}},\
  }\bibfield  {title} {\bibinfo {title} {{Analysis of Charge Order in the
  Kagome Metal $A{\mathrm{V}}_{3}{\mathrm{Sb}}_{5}$
  ($A=\mathrm{K},\mathrm{Rb},\mathrm{Cs}$)}},\ }\href
  {https://doi.org/10.1103/PhysRevLett.127.217601} {\bibfield  {journal}
  {\bibinfo  {journal} {Phys. Rev. Lett.}\ }\textbf {\bibinfo {volume} {127}},\
  \bibinfo {pages} {217601} (\bibinfo {year} {2021})}\BibitemShut {NoStop}%
\bibitem [{\citenamefont {Lin}\ and\ \citenamefont
  {Nandkishore}(2021)}]{lin21prb}%
  \BibitemOpen
  \bibfield  {author} {\bibinfo {author} {\bibfnamefont {Y.-P.}\ \bibnamefont
  {Lin}}\ and\ \bibinfo {author} {\bibfnamefont {R.~M.}\ \bibnamefont
  {Nandkishore}},\ }\bibfield  {title} {\bibinfo {title} {{Complex charge
  density waves at Van Hove singularity on hexagonal lattices: Haldane-model
  phase diagram and potential realization in the kagome metals
  $A{V}_{3}{\mathrm{Sb}}_{5}$ ($A$=K, Rb, Cs)}},\ }\href
  {https://doi.org/10.1103/PhysRevB.104.045122} {\bibfield  {journal} {\bibinfo
   {journal} {Phys. Rev. B}\ }\textbf {\bibinfo {volume} {104}},\ \bibinfo
  {pages} {045122} (\bibinfo {year} {2021})}\BibitemShut {NoStop}%
\bibitem [{\citenamefont {Park}\ \emph {et~al.}(2021)\citenamefont {Park},
  \citenamefont {Ye},\ and\ \citenamefont {Balents}}]{park21prb}%
  \BibitemOpen
  \bibfield  {author} {\bibinfo {author} {\bibfnamefont {T.}~\bibnamefont
  {Park}}, \bibinfo {author} {\bibfnamefont {M.}~\bibnamefont {Ye}},\ and\
  \bibinfo {author} {\bibfnamefont {L.}~\bibnamefont {Balents}},\ }\bibfield
  {title} {\bibinfo {title} {Electronic instabilities of kagome metals: Saddle
  points and landau theory},\ }\href
  {https://doi.org/10.1103/PhysRevB.104.035142} {\bibfield  {journal} {\bibinfo
   {journal} {Phys. Rev. B}\ }\textbf {\bibinfo {volume} {104}},\ \bibinfo
  {pages} {035142} (\bibinfo {year} {2021})}\BibitemShut {NoStop}%
\bibitem [{\citenamefont {{Setty}}\ \emph {et~al.}(2021)\citenamefont
  {{Setty}}, \citenamefont {{Hu}}, \citenamefont {{Chen}},\ and\ \citenamefont
  {{Si}}}]{setty21ax}%
  \BibitemOpen
  \bibfield  {author} {\bibinfo {author} {\bibfnamefont {C.}~\bibnamefont
  {{Setty}}}, \bibinfo {author} {\bibfnamefont {H.}~\bibnamefont {{Hu}}},
  \bibinfo {author} {\bibfnamefont {L.}~\bibnamefont {{Chen}}},\ and\ \bibinfo
  {author} {\bibfnamefont {Q.}~\bibnamefont {{Si}}},\ }\bibfield  {title}
  {\bibinfo {title} {{Electron correlations and $T$-breaking density wave order
  in a $\mathbb{Z}_2$ kagome metal}},\ }\href@noop {} {\bibfield  {journal}
  {\bibinfo  {journal} {arXiv e-prints}\ ,\ \bibinfo {eid} {arXiv:2105.15204}}
  (\bibinfo {year} {2021})},\ \Eprint {https://arxiv.org/abs/2105.15204}
  {arXiv:2105.15204 [cond-mat.str-el]} \BibitemShut {NoStop}%
\bibitem [{\citenamefont {Feng}\ \emph
  {et~al.}(2021{\natexlab{b}})\citenamefont {Feng}, \citenamefont {Zhang},
  \citenamefont {Jiang},\ and\ \citenamefont {Hu}}]{feng21prb}%
  \BibitemOpen
  \bibfield  {author} {\bibinfo {author} {\bibfnamefont {X.}~\bibnamefont
  {Feng}}, \bibinfo {author} {\bibfnamefont {Y.}~\bibnamefont {Zhang}},
  \bibinfo {author} {\bibfnamefont {K.}~\bibnamefont {Jiang}},\ and\ \bibinfo
  {author} {\bibfnamefont {J.}~\bibnamefont {Hu}},\ }\bibfield  {title}
  {\bibinfo {title} {Low-energy effective theory and symmetry classification of
  flux phases on the kagome lattice},\ }\href
  {https://doi.org/10.1103/PhysRevB.104.165136} {\bibfield  {journal} {\bibinfo
   {journal} {Phys. Rev. B}\ }\textbf {\bibinfo {volume} {104}},\ \bibinfo
  {pages} {165136} (\bibinfo {year} {2021}{\natexlab{b}})}\BibitemShut
  {NoStop}%
\bibitem [{\citenamefont {Miao}\ \emph {et~al.}(2021)\citenamefont {Miao},
  \citenamefont {Li}, \citenamefont {Meier}, \citenamefont {Huon},
  \citenamefont {Lee}, \citenamefont {Said}, \citenamefont {Lei}, \citenamefont
  {Ortiz}, \citenamefont {Wilson}, \citenamefont {Yin}, \citenamefont {Hasan},
  \citenamefont {Wang}, \citenamefont {Tan},\ and\ \citenamefont
  {Yan}}]{miao21prb}%
  \BibitemOpen
  \bibfield  {author} {\bibinfo {author} {\bibfnamefont {H.}~\bibnamefont
  {Miao}}, \bibinfo {author} {\bibfnamefont {H.~X.}\ \bibnamefont {Li}},
  \bibinfo {author} {\bibfnamefont {W.~R.}\ \bibnamefont {Meier}}, \bibinfo
  {author} {\bibfnamefont {A.}~\bibnamefont {Huon}}, \bibinfo {author}
  {\bibfnamefont {H.~N.}\ \bibnamefont {Lee}}, \bibinfo {author} {\bibfnamefont
  {A.}~\bibnamefont {Said}}, \bibinfo {author} {\bibfnamefont {H.~C.}\
  \bibnamefont {Lei}}, \bibinfo {author} {\bibfnamefont {B.~R.}\ \bibnamefont
  {Ortiz}}, \bibinfo {author} {\bibfnamefont {S.~D.}\ \bibnamefont {Wilson}},
  \bibinfo {author} {\bibfnamefont {J.~X.}\ \bibnamefont {Yin}}, \bibinfo
  {author} {\bibfnamefont {M.~Z.}\ \bibnamefont {Hasan}}, \bibinfo {author}
  {\bibfnamefont {Z.}~\bibnamefont {Wang}}, \bibinfo {author} {\bibfnamefont
  {H.}~\bibnamefont {Tan}},\ and\ \bibinfo {author} {\bibfnamefont
  {B.}~\bibnamefont {Yan}},\ }\bibfield  {title} {\bibinfo {title} {{Geometry
  of the charge density wave in the kagome metal
  $A{\mathrm{V}}_{3}{\mathrm{Sb}}_{5}$}},\ }\href
  {https://doi.org/10.1103/PhysRevB.104.195132} {\bibfield  {journal} {\bibinfo
   {journal} {Phys. Rev. B}\ }\textbf {\bibinfo {volume} {104}},\ \bibinfo
  {pages} {195132} (\bibinfo {year} {2021})}\BibitemShut {NoStop}%
\bibitem [{\citenamefont {Christensen}\ \emph {et~al.}(2021)\citenamefont
  {Christensen}, \citenamefont {Birol}, \citenamefont {Andersen},\ and\
  \citenamefont {Fernandes}}]{christensen21prb}%
  \BibitemOpen
  \bibfield  {author} {\bibinfo {author} {\bibfnamefont {M.~H.}\ \bibnamefont
  {Christensen}}, \bibinfo {author} {\bibfnamefont {T.}~\bibnamefont {Birol}},
  \bibinfo {author} {\bibfnamefont {B.~M.}\ \bibnamefont {Andersen}},\ and\
  \bibinfo {author} {\bibfnamefont {R.~M.}\ \bibnamefont {Fernandes}},\
  }\bibfield  {title} {\bibinfo {title} {{Theory of the charge density wave in
  $A{\mathrm{V}}_{3}{\mathrm{Sb}}_{5}$ kagome metals}},\ }\href
  {https://doi.org/10.1103/PhysRevB.104.214513} {\bibfield  {journal} {\bibinfo
   {journal} {Phys. Rev. B}\ }\textbf {\bibinfo {volume} {104}},\ \bibinfo
  {pages} {214513} (\bibinfo {year} {2021})}\BibitemShut {NoStop}%
\bibitem [{\citenamefont {{Lin}}(2021)}]{lin21ax}%
  \BibitemOpen
  \bibfield  {author} {\bibinfo {author} {\bibfnamefont {Y.-P.}\ \bibnamefont
  {{Lin}}},\ }\bibfield  {title} {\bibinfo {title} {{Higher-order topological
  insulators from $3Q$ charge bond orders on hexagonal lattices: A hint to
  kagome metals}},\ }\href@noop {} {\bibfield  {journal} {\bibinfo  {journal}
  {arXiv e-prints}\ ,\ \bibinfo {eid} {arXiv:2106.09717}} (\bibinfo {year}
  {2021})},\ \Eprint {https://arxiv.org/abs/2106.09717} {arXiv:2106.09717
  [cond-mat.str-el]} \BibitemShut {NoStop}%
\bibitem [{\citenamefont {Affleck}\ and\ \citenamefont
  {Marston}(1988)}]{affleck88prb}%
  \BibitemOpen
  \bibfield  {author} {\bibinfo {author} {\bibfnamefont {I.}~\bibnamefont
  {Affleck}}\ and\ \bibinfo {author} {\bibfnamefont {J.~B.}\ \bibnamefont
  {Marston}},\ }\bibfield  {title} {\bibinfo {title} {Large-n limit of the
  heisenberg-hubbard model: Implications for high-${T}_{c}$ superconductors},\
  }\href {https://doi.org/10.1103/PhysRevB.37.3774} {\bibfield  {journal}
  {\bibinfo  {journal} {Phys. Rev. B}\ }\textbf {\bibinfo {volume} {37}},\
  \bibinfo {pages} {3774} (\bibinfo {year} {1988})}\BibitemShut {NoStop}%
\bibitem [{\citenamefont {Varma}(1997)}]{varma}%
  \BibitemOpen
  \bibfield  {author} {\bibinfo {author} {\bibfnamefont {C.~M.}\ \bibnamefont
  {Varma}},\ }\bibfield  {title} {\bibinfo {title} {Non-fermi-liquid states and
  pairing instability of a general model of copper oxide metals},\ }\href
  {https://doi.org/10.1103/PhysRevB.55.14554} {\bibfield  {journal} {\bibinfo
  {journal} {Phys. Rev. B}\ }\textbf {\bibinfo {volume} {55}},\ \bibinfo
  {pages} {14554} (\bibinfo {year} {1997})}\BibitemShut {NoStop}%
\bibitem [{\citenamefont {Nayak}(2000)}]{nayak00prb}%
  \BibitemOpen
  \bibfield  {author} {\bibinfo {author} {\bibfnamefont {C.}~\bibnamefont
  {Nayak}},\ }\bibfield  {title} {\bibinfo {title} {Density-wave states of
  nonzero angular momentum},\ }\href {https://doi.org/10.1103/PhysRevB.62.4880}
  {\bibfield  {journal} {\bibinfo  {journal} {Phys. Rev. B}\ }\textbf {\bibinfo
  {volume} {62}},\ \bibinfo {pages} {4880} (\bibinfo {year}
  {2000})}\BibitemShut {NoStop}%
\bibitem [{\citenamefont {Chakravarty}\ \emph {et~al.}(2001)\citenamefont
  {Chakravarty}, \citenamefont {Laughlin}, \citenamefont {Morr},\ and\
  \citenamefont {Nayak}}]{chakravarty01prb}%
  \BibitemOpen
  \bibfield  {author} {\bibinfo {author} {\bibfnamefont {S.}~\bibnamefont
  {Chakravarty}}, \bibinfo {author} {\bibfnamefont {R.~B.}\ \bibnamefont
  {Laughlin}}, \bibinfo {author} {\bibfnamefont {D.~K.}\ \bibnamefont {Morr}},\
  and\ \bibinfo {author} {\bibfnamefont {C.}~\bibnamefont {Nayak}},\ }\bibfield
   {title} {\bibinfo {title} {Hidden order in the cuprates},\ }\href
  {https://doi.org/10.1103/PhysRevB.63.094503} {\bibfield  {journal} {\bibinfo
  {journal} {Phys. Rev. B}\ }\textbf {\bibinfo {volume} {63}},\ \bibinfo
  {pages} {094503} (\bibinfo {year} {2001})}\BibitemShut {NoStop}%
\bibitem [{\citenamefont {Venderbos}(2016)}]{venderbos16prb}%
  \BibitemOpen
  \bibfield  {author} {\bibinfo {author} {\bibfnamefont {J.~W.~F.}\
  \bibnamefont {Venderbos}},\ }\bibfield  {title} {\bibinfo {title} {Symmetry
  analysis of translational symmetry broken density waves: Application to
  hexagonal lattices in two dimensions},\ }\href
  {https://doi.org/10.1103/PhysRevB.93.115107} {\bibfield  {journal} {\bibinfo
  {journal} {Phys. Rev. B}\ }\textbf {\bibinfo {volume} {93}},\ \bibinfo
  {pages} {115107} (\bibinfo {year} {2016})}\BibitemShut {NoStop}%
\bibitem [{\citenamefont {Lin}\ and\ \citenamefont
  {Nandkishore}(2019)}]{lin19prb}%
  \BibitemOpen
  \bibfield  {author} {\bibinfo {author} {\bibfnamefont {Y.-P.}\ \bibnamefont
  {Lin}}\ and\ \bibinfo {author} {\bibfnamefont {R.~M.}\ \bibnamefont
  {Nandkishore}},\ }\bibfield  {title} {\bibinfo {title} {Chiral twist on the
  high-${T}_{c}$ phase diagram in moir\'e heterostructures},\ }\href
  {https://doi.org/10.1103/PhysRevB.100.085136} {\bibfield  {journal} {\bibinfo
   {journal} {Phys. Rev. B}\ }\textbf {\bibinfo {volume} {100}},\ \bibinfo
  {pages} {085136} (\bibinfo {year} {2019})}\BibitemShut {NoStop}%
\bibitem [{\citenamefont {Ortiz}\ \emph {et~al.}(2021)\citenamefont {Ortiz},
  \citenamefont {Sarte}, \citenamefont {Kenney}, \citenamefont {Graf},
  \citenamefont {Teicher}, \citenamefont {Seshadri},\ and\ \citenamefont
  {Wilson}}]{ortiz21prm}%
  \BibitemOpen
  \bibfield  {author} {\bibinfo {author} {\bibfnamefont {B.~R.}\ \bibnamefont
  {Ortiz}}, \bibinfo {author} {\bibfnamefont {P.~M.}\ \bibnamefont {Sarte}},
  \bibinfo {author} {\bibfnamefont {E.~M.}\ \bibnamefont {Kenney}}, \bibinfo
  {author} {\bibfnamefont {M.~J.}\ \bibnamefont {Graf}}, \bibinfo {author}
  {\bibfnamefont {S.~M.~L.}\ \bibnamefont {Teicher}}, \bibinfo {author}
  {\bibfnamefont {R.}~\bibnamefont {Seshadri}},\ and\ \bibinfo {author}
  {\bibfnamefont {S.~D.}\ \bibnamefont {Wilson}},\ }\bibfield  {title}
  {\bibinfo {title} {{Superconductivity in the ${\mathbb{Z}}_{2}$ kagome metal
  ${\mathrm{KV}}_{3}{\mathrm{Sb}}_{5}$}},\ }\href
  {https://doi.org/10.1103/PhysRevMaterials.5.034801} {\bibfield  {journal}
  {\bibinfo  {journal} {Phys. Rev. Materials}\ }\textbf {\bibinfo {volume}
  {5}},\ \bibinfo {pages} {034801} (\bibinfo {year} {2021})}\BibitemShut
  {NoStop}%
\bibitem [{\citenamefont {{Wang}}\ \emph {et~al.}(2020)\citenamefont {{Wang}},
  \citenamefont {{Yang}}, \citenamefont {{Sivakumar}}, \citenamefont {{Ortiz}},
  \citenamefont {{Teicher}}, \citenamefont {{Wu}}, \citenamefont
  {{Srivastava}}, \citenamefont {{Garg}}, \citenamefont {{Liu}}, \citenamefont
  {{Parkin}}, \citenamefont {{Toberer}}, \citenamefont {{McQueen}},
  \citenamefont {{Wilson}},\ and\ \citenamefont {{Ali}}}]{wang20ax}%
  \BibitemOpen
  \bibfield  {author} {\bibinfo {author} {\bibfnamefont {Y.}~\bibnamefont
  {{Wang}}}, \bibinfo {author} {\bibfnamefont {S.}~\bibnamefont {{Yang}}},
  \bibinfo {author} {\bibfnamefont {P.~K.}\ \bibnamefont {{Sivakumar}}},
  \bibinfo {author} {\bibfnamefont {B.~R.}\ \bibnamefont {{Ortiz}}}, \bibinfo
  {author} {\bibfnamefont {S.~M.~L.}\ \bibnamefont {{Teicher}}}, \bibinfo
  {author} {\bibfnamefont {H.}~\bibnamefont {{Wu}}}, \bibinfo {author}
  {\bibfnamefont {A.~K.}\ \bibnamefont {{Srivastava}}}, \bibinfo {author}
  {\bibfnamefont {C.}~\bibnamefont {{Garg}}}, \bibinfo {author} {\bibfnamefont
  {D.}~\bibnamefont {{Liu}}}, \bibinfo {author} {\bibfnamefont {S.~S.~P.}\
  \bibnamefont {{Parkin}}}, \bibinfo {author} {\bibfnamefont {E.~S.}\
  \bibnamefont {{Toberer}}}, \bibinfo {author} {\bibfnamefont {T.}~\bibnamefont
  {{McQueen}}}, \bibinfo {author} {\bibfnamefont {S.~D.}\ \bibnamefont
  {{Wilson}}},\ and\ \bibinfo {author} {\bibfnamefont {M.~N.}\ \bibnamefont
  {{Ali}}},\ }\bibfield  {title} {\bibinfo {title} {{Proximity-induced
  spin-triplet superconductivity and edge supercurrent in the topological
  Kagome metal, $\mathrm{K_{1-x}V_3Sb_5}$}},\ }\href@noop {} {\bibfield
  {journal} {\bibinfo  {journal} {arXiv e-prints}\ ,\ \bibinfo {eid}
  {arXiv:2012.05898}} (\bibinfo {year} {2020})},\ \Eprint
  {https://arxiv.org/abs/2012.05898} {arXiv:2012.05898 [cond-mat.supr-con]}
  \BibitemShut {NoStop}%
\bibitem [{\citenamefont {{Zhao}}\ \emph
  {et~al.}(2021{\natexlab{b}})\citenamefont {{Zhao}}, \citenamefont {{Wang}},
  \citenamefont {{Xia}}, \citenamefont {{Yin}}, \citenamefont {{Ni}},
  \citenamefont {{Huang}}, \citenamefont {{Tu}}, \citenamefont {{Tao}},
  \citenamefont {{Tu}}, \citenamefont {{Gong}}, \citenamefont {{Lei}},
  \citenamefont {{Guo}}, \citenamefont {{Yang}},\ and\ \citenamefont
  {{Li}}}]{zhao21axsc}%
  \BibitemOpen
  \bibfield  {author} {\bibinfo {author} {\bibfnamefont {C.~C.}\ \bibnamefont
  {{Zhao}}}, \bibinfo {author} {\bibfnamefont {L.~S.}\ \bibnamefont {{Wang}}},
  \bibinfo {author} {\bibfnamefont {W.}~\bibnamefont {{Xia}}}, \bibinfo
  {author} {\bibfnamefont {Q.~W.}\ \bibnamefont {{Yin}}}, \bibinfo {author}
  {\bibfnamefont {J.~M.}\ \bibnamefont {{Ni}}}, \bibinfo {author}
  {\bibfnamefont {Y.~Y.}\ \bibnamefont {{Huang}}}, \bibinfo {author}
  {\bibfnamefont {C.~P.}\ \bibnamefont {{Tu}}}, \bibinfo {author}
  {\bibfnamefont {Z.~C.}\ \bibnamefont {{Tao}}}, \bibinfo {author}
  {\bibfnamefont {Z.~J.}\ \bibnamefont {{Tu}}}, \bibinfo {author}
  {\bibfnamefont {C.~S.}\ \bibnamefont {{Gong}}}, \bibinfo {author}
  {\bibfnamefont {H.~C.}\ \bibnamefont {{Lei}}}, \bibinfo {author}
  {\bibfnamefont {Y.~F.}\ \bibnamefont {{Guo}}}, \bibinfo {author}
  {\bibfnamefont {X.~F.}\ \bibnamefont {{Yang}}},\ and\ \bibinfo {author}
  {\bibfnamefont {S.~Y.}\ \bibnamefont {{Li}}},\ }\bibfield  {title} {\bibinfo
  {title} {{Nodal superconductivity and superconducting domes in the
  topological Kagome metal CsV3Sb5}},\ }\href@noop {} {\bibfield  {journal}
  {\bibinfo  {journal} {arXiv e-prints}\ ,\ \bibinfo {eid} {arXiv:2102.08356}}
  (\bibinfo {year} {2021}{\natexlab{b}})},\ \Eprint
  {https://arxiv.org/abs/2102.08356} {arXiv:2102.08356 [cond-mat.supr-con]}
  \BibitemShut {NoStop}%
\bibitem [{\citenamefont {{Duan}}\ \emph {et~al.}(2021)\citenamefont {{Duan}},
  \citenamefont {{Nie}}, \citenamefont {{Luo}}, \citenamefont {{Yu}},
  \citenamefont {{Ortiz}}, \citenamefont {{Yin}}, \citenamefont {{Su}},
  \citenamefont {{Du}}, \citenamefont {{Wang}}, \citenamefont {{Chen}},
  \citenamefont {{Lu}}, \citenamefont {{Ying}}, \citenamefont {{Wilson}},
  \citenamefont {{Chen}}, \citenamefont {{Song}},\ and\ \citenamefont
  {{Yuan}}}]{duan21scp}%
  \BibitemOpen
  \bibfield  {author} {\bibinfo {author} {\bibfnamefont {W.}~\bibnamefont
  {{Duan}}}, \bibinfo {author} {\bibfnamefont {Z.}~\bibnamefont {{Nie}}},
  \bibinfo {author} {\bibfnamefont {S.}~\bibnamefont {{Luo}}}, \bibinfo
  {author} {\bibfnamefont {F.}~\bibnamefont {{Yu}}}, \bibinfo {author}
  {\bibfnamefont {B.~R.}\ \bibnamefont {{Ortiz}}}, \bibinfo {author}
  {\bibfnamefont {L.}~\bibnamefont {{Yin}}}, \bibinfo {author} {\bibfnamefont
  {H.}~\bibnamefont {{Su}}}, \bibinfo {author} {\bibfnamefont {F.}~\bibnamefont
  {{Du}}}, \bibinfo {author} {\bibfnamefont {A.}~\bibnamefont {{Wang}}},
  \bibinfo {author} {\bibfnamefont {Y.}~\bibnamefont {{Chen}}}, \bibinfo
  {author} {\bibfnamefont {X.}~\bibnamefont {{Lu}}}, \bibinfo {author}
  {\bibfnamefont {J.}~\bibnamefont {{Ying}}}, \bibinfo {author} {\bibfnamefont
  {S.~D.}\ \bibnamefont {{Wilson}}}, \bibinfo {author} {\bibfnamefont
  {X.}~\bibnamefont {{Chen}}}, \bibinfo {author} {\bibfnamefont
  {Y.}~\bibnamefont {{Song}}},\ and\ \bibinfo {author} {\bibfnamefont
  {H.}~\bibnamefont {{Yuan}}},\ }\bibfield  {title} {\bibinfo {title}
  {{Nodeless superconductivity in the kagome metal CsV$_{3}$Sb$_{5}$}},\ }\href
  {https://doi.org/10.1007/s11433-021-1747-7} {\bibfield  {journal} {\bibinfo
  {journal} {Sci. China-Phys., Mech. Astron.}\ }\textbf {\bibinfo {volume}
  {64}},\ \bibinfo {pages} {107462} (\bibinfo {year} {2021})}\BibitemShut
  {NoStop}%
\bibitem [{\citenamefont {Zhang}\ \emph {et~al.}(2021)\citenamefont {Zhang},
  \citenamefont {Chen}, \citenamefont {Zhou}, \citenamefont {Yuan},
  \citenamefont {Wang}, \citenamefont {Wang}, \citenamefont {Yang},
  \citenamefont {An}, \citenamefont {Zhang}, \citenamefont {Zhu}, \citenamefont
  {Zhou}, \citenamefont {Chen}, \citenamefont {Zhou},\ and\ \citenamefont
  {Yang}}]{zhang21prb}%
  \BibitemOpen
  \bibfield  {author} {\bibinfo {author} {\bibfnamefont {Z.}~\bibnamefont
  {Zhang}}, \bibinfo {author} {\bibfnamefont {Z.}~\bibnamefont {Chen}},
  \bibinfo {author} {\bibfnamefont {Y.}~\bibnamefont {Zhou}}, \bibinfo {author}
  {\bibfnamefont {Y.}~\bibnamefont {Yuan}}, \bibinfo {author} {\bibfnamefont
  {S.}~\bibnamefont {Wang}}, \bibinfo {author} {\bibfnamefont {J.}~\bibnamefont
  {Wang}}, \bibinfo {author} {\bibfnamefont {H.}~\bibnamefont {Yang}}, \bibinfo
  {author} {\bibfnamefont {C.}~\bibnamefont {An}}, \bibinfo {author}
  {\bibfnamefont {L.}~\bibnamefont {Zhang}}, \bibinfo {author} {\bibfnamefont
  {X.}~\bibnamefont {Zhu}}, \bibinfo {author} {\bibfnamefont {Y.}~\bibnamefont
  {Zhou}}, \bibinfo {author} {\bibfnamefont {X.}~\bibnamefont {Chen}}, \bibinfo
  {author} {\bibfnamefont {J.}~\bibnamefont {Zhou}},\ and\ \bibinfo {author}
  {\bibfnamefont {Z.}~\bibnamefont {Yang}},\ }\bibfield  {title} {\bibinfo
  {title} {{Pressure-induced reemergence of superconductivity in the
  topological kagome metal $\mathrm{Cs}{\mathrm{V}}_{3}{\mathrm{Sb}}_{5}$}},\
  }\href {https://doi.org/10.1103/PhysRevB.103.224513} {\bibfield  {journal}
  {\bibinfo  {journal} {Phys. Rev. B}\ }\textbf {\bibinfo {volume} {103}},\
  \bibinfo {pages} {224513} (\bibinfo {year} {2021})}\BibitemShut {NoStop}%
\bibitem [{\citenamefont {Ni}\ \emph {et~al.}(2021)\citenamefont {Ni},
  \citenamefont {Ma}, \citenamefont {Zhang}, \citenamefont {Yuan},
  \citenamefont {Yang}, \citenamefont {Lu}, \citenamefont {Wang}, \citenamefont
  {Sun}, \citenamefont {Zhao}, \citenamefont {Li}, \citenamefont {Liu},
  \citenamefont {Zhang}, \citenamefont {Chen}, \citenamefont {Jin},
  \citenamefont {Cheng}, \citenamefont {Yu}, \citenamefont {Zhou},
  \citenamefont {Dong}, \citenamefont {Hu}, \citenamefont {Gao},\ and\
  \citenamefont {Zhao}}]{ni21cpl}%
  \BibitemOpen
  \bibfield  {author} {\bibinfo {author} {\bibfnamefont {S.}~\bibnamefont
  {Ni}}, \bibinfo {author} {\bibfnamefont {S.}~\bibnamefont {Ma}}, \bibinfo
  {author} {\bibfnamefont {Y.}~\bibnamefont {Zhang}}, \bibinfo {author}
  {\bibfnamefont {J.}~\bibnamefont {Yuan}}, \bibinfo {author} {\bibfnamefont
  {H.}~\bibnamefont {Yang}}, \bibinfo {author} {\bibfnamefont {Z.}~\bibnamefont
  {Lu}}, \bibinfo {author} {\bibfnamefont {N.}~\bibnamefont {Wang}}, \bibinfo
  {author} {\bibfnamefont {J.}~\bibnamefont {Sun}}, \bibinfo {author}
  {\bibfnamefont {Z.}~\bibnamefont {Zhao}}, \bibinfo {author} {\bibfnamefont
  {D.}~\bibnamefont {Li}}, \bibinfo {author} {\bibfnamefont {S.}~\bibnamefont
  {Liu}}, \bibinfo {author} {\bibfnamefont {H.}~\bibnamefont {Zhang}}, \bibinfo
  {author} {\bibfnamefont {H.}~\bibnamefont {Chen}}, \bibinfo {author}
  {\bibfnamefont {K.}~\bibnamefont {Jin}}, \bibinfo {author} {\bibfnamefont
  {J.}~\bibnamefont {Cheng}}, \bibinfo {author} {\bibfnamefont
  {L.}~\bibnamefont {Yu}}, \bibinfo {author} {\bibfnamefont {F.}~\bibnamefont
  {Zhou}}, \bibinfo {author} {\bibfnamefont {X.}~\bibnamefont {Dong}}, \bibinfo
  {author} {\bibfnamefont {J.}~\bibnamefont {Hu}}, \bibinfo {author}
  {\bibfnamefont {H.-J.}\ \bibnamefont {Gao}},\ and\ \bibinfo {author}
  {\bibfnamefont {Z.}~\bibnamefont {Zhao}},\ }\bibfield  {title} {\bibinfo
  {title} {{Anisotropic Superconducting Properties of Kagome Metal
  {CsV}3Sb5}},\ }\href {https://doi.org/10.1088/0256-307x/38/5/057403}
  {\bibfield  {journal} {\bibinfo  {journal} {Chin. Phys. Lett.}\ }\textbf
  {\bibinfo {volume} {38}},\ \bibinfo {pages} {057403} (\bibinfo {year}
  {2021})}\BibitemShut {NoStop}%
\bibitem [{\citenamefont {{Xiang}}\ \emph {et~al.}(2021)\citenamefont
  {{Xiang}}, \citenamefont {{Li}}, \citenamefont {{Li}}, \citenamefont {{Xie}},
  \citenamefont {{Yang}}, \citenamefont {{Wang}}, \citenamefont {{Yao}},\ and\
  \citenamefont {{Wen}}}]{xiang21nc}%
  \BibitemOpen
  \bibfield  {author} {\bibinfo {author} {\bibfnamefont {Y.}~\bibnamefont
  {{Xiang}}}, \bibinfo {author} {\bibfnamefont {Q.}~\bibnamefont {{Li}}},
  \bibinfo {author} {\bibfnamefont {Y.}~\bibnamefont {{Li}}}, \bibinfo {author}
  {\bibfnamefont {W.}~\bibnamefont {{Xie}}}, \bibinfo {author} {\bibfnamefont
  {H.}~\bibnamefont {{Yang}}}, \bibinfo {author} {\bibfnamefont
  {Z.}~\bibnamefont {{Wang}}}, \bibinfo {author} {\bibfnamefont
  {Y.}~\bibnamefont {{Yao}}},\ and\ \bibinfo {author} {\bibfnamefont {H.-H.}\
  \bibnamefont {{Wen}}},\ }\bibfield  {title} {\bibinfo {title} {{Twofold
  symmetry of c-axis resistivity in topological kagome superconductor
  CsV$_{3}$Sb$_{5}$ with in-plane rotating magnetic field}},\ }\href
  {https://doi.org/10.1038/s41467-021-27084-z} {\bibfield  {journal} {\bibinfo
  {journal} {Nat. Commun.}\ }\textbf {\bibinfo {volume} {12}},\ \bibinfo
  {pages} {6727} (\bibinfo {year} {2021})}\BibitemShut {NoStop}%
\bibitem [{\citenamefont {Xu}\ \emph {et~al.}(2021)\citenamefont {Xu},
  \citenamefont {Yan}, \citenamefont {Yin}, \citenamefont {Xia}, \citenamefont
  {Fang}, \citenamefont {Chen}, \citenamefont {Li}, \citenamefont {Yang},
  \citenamefont {Guo},\ and\ \citenamefont {Feng}}]{xu21prl}%
  \BibitemOpen
  \bibfield  {author} {\bibinfo {author} {\bibfnamefont {H.-S.}\ \bibnamefont
  {Xu}}, \bibinfo {author} {\bibfnamefont {Y.-J.}\ \bibnamefont {Yan}},
  \bibinfo {author} {\bibfnamefont {R.}~\bibnamefont {Yin}}, \bibinfo {author}
  {\bibfnamefont {W.}~\bibnamefont {Xia}}, \bibinfo {author} {\bibfnamefont
  {S.}~\bibnamefont {Fang}}, \bibinfo {author} {\bibfnamefont {Z.}~\bibnamefont
  {Chen}}, \bibinfo {author} {\bibfnamefont {Y.}~\bibnamefont {Li}}, \bibinfo
  {author} {\bibfnamefont {W.}~\bibnamefont {Yang}}, \bibinfo {author}
  {\bibfnamefont {Y.}~\bibnamefont {Guo}},\ and\ \bibinfo {author}
  {\bibfnamefont {D.-L.}\ \bibnamefont {Feng}},\ }\bibfield  {title} {\bibinfo
  {title} {{Multiband Superconductivity with Sign-Preserving Order Parameter in
  Kagome Superconductor ${\mathrm{CsV}}_{3}{\mathrm{Sb}}_{5}$}},\ }\href
  {https://doi.org/10.1103/PhysRevLett.127.187004} {\bibfield  {journal}
  {\bibinfo  {journal} {Phys. Rev. Lett.}\ }\textbf {\bibinfo {volume} {127}},\
  \bibinfo {pages} {187004} (\bibinfo {year} {2021})}\BibitemShut {NoStop}%
\bibitem [{\citenamefont {Zhu}\ \emph {et~al.}(2022)\citenamefont {Zhu},
  \citenamefont {Yang}, \citenamefont {Xia}, \citenamefont {Yin}, \citenamefont
  {Wang}, \citenamefont {Zhao}, \citenamefont {Dai}, \citenamefont {Tu},
  \citenamefont {Song}, \citenamefont {Tao}, \citenamefont {Tu}, \citenamefont
  {Gong}, \citenamefont {Lei}, \citenamefont {Guo},\ and\ \citenamefont
  {Li}}]{zhu22prb}%
  \BibitemOpen
  \bibfield  {author} {\bibinfo {author} {\bibfnamefont {C.~C.}\ \bibnamefont
  {Zhu}}, \bibinfo {author} {\bibfnamefont {X.~F.}\ \bibnamefont {Yang}},
  \bibinfo {author} {\bibfnamefont {W.}~\bibnamefont {Xia}}, \bibinfo {author}
  {\bibfnamefont {Q.~W.}\ \bibnamefont {Yin}}, \bibinfo {author} {\bibfnamefont
  {L.~S.}\ \bibnamefont {Wang}}, \bibinfo {author} {\bibfnamefont {C.~C.}\
  \bibnamefont {Zhao}}, \bibinfo {author} {\bibfnamefont {D.~Z.}\ \bibnamefont
  {Dai}}, \bibinfo {author} {\bibfnamefont {C.~P.}\ \bibnamefont {Tu}},
  \bibinfo {author} {\bibfnamefont {B.~Q.}\ \bibnamefont {Song}}, \bibinfo
  {author} {\bibfnamefont {Z.~C.}\ \bibnamefont {Tao}}, \bibinfo {author}
  {\bibfnamefont {Z.~J.}\ \bibnamefont {Tu}}, \bibinfo {author} {\bibfnamefont
  {C.~S.}\ \bibnamefont {Gong}}, \bibinfo {author} {\bibfnamefont {H.~C.}\
  \bibnamefont {Lei}}, \bibinfo {author} {\bibfnamefont {Y.~F.}\ \bibnamefont
  {Guo}},\ and\ \bibinfo {author} {\bibfnamefont {S.~Y.}\ \bibnamefont {Li}},\
  }\bibfield  {title} {\bibinfo {title} {{Double-dome superconductivity under
  pressure in the V-based kagome metals $A{\mathrm{V}}_{3}{\mathrm{Sb}}_{5}$
  ($A=\mathrm{Rb}$ and K)}},\ }\href
  {https://doi.org/10.1103/PhysRevB.105.094507} {\bibfield  {journal} {\bibinfo
   {journal} {Phys. Rev. B}\ }\textbf {\bibinfo {volume} {105}},\ \bibinfo
  {pages} {094507} (\bibinfo {year} {2022})}\BibitemShut {NoStop}%
\bibitem [{\citenamefont {Wu}\ \emph {et~al.}(2021)\citenamefont {Wu},
  \citenamefont {Schwemmer}, \citenamefont {M\"uller}, \citenamefont
  {Consiglio}, \citenamefont {Sangiovanni}, \citenamefont {Di~Sante},
  \citenamefont {Iqbal}, \citenamefont {Hanke}, \citenamefont {Schnyder},
  \citenamefont {Denner}, \citenamefont {Fischer}, \citenamefont {Neupert},\
  and\ \citenamefont {Thomale}}]{wu21prl}%
  \BibitemOpen
  \bibfield  {author} {\bibinfo {author} {\bibfnamefont {X.}~\bibnamefont
  {Wu}}, \bibinfo {author} {\bibfnamefont {T.}~\bibnamefont {Schwemmer}},
  \bibinfo {author} {\bibfnamefont {T.}~\bibnamefont {M\"uller}}, \bibinfo
  {author} {\bibfnamefont {A.}~\bibnamefont {Consiglio}}, \bibinfo {author}
  {\bibfnamefont {G.}~\bibnamefont {Sangiovanni}}, \bibinfo {author}
  {\bibfnamefont {D.}~\bibnamefont {Di~Sante}}, \bibinfo {author}
  {\bibfnamefont {Y.}~\bibnamefont {Iqbal}}, \bibinfo {author} {\bibfnamefont
  {W.}~\bibnamefont {Hanke}}, \bibinfo {author} {\bibfnamefont {A.~P.}\
  \bibnamefont {Schnyder}}, \bibinfo {author} {\bibfnamefont {M.~M.}\
  \bibnamefont {Denner}}, \bibinfo {author} {\bibfnamefont {M.~H.}\
  \bibnamefont {Fischer}}, \bibinfo {author} {\bibfnamefont {T.}~\bibnamefont
  {Neupert}},\ and\ \bibinfo {author} {\bibfnamefont {R.}~\bibnamefont
  {Thomale}},\ }\bibfield  {title} {\bibinfo {title} {{Nature of Unconventional
  Pairing in the Kagome Superconductors $A{\mathrm{V}}_{3}{\mathrm{Sb}}_{5}$
  ($A=\mathrm{K},\mathrm{Rb},\mathrm{Cs}$)}},\ }\href
  {https://doi.org/10.1103/PhysRevLett.127.177001} {\bibfield  {journal}
  {\bibinfo  {journal} {Phys. Rev. Lett.}\ }\textbf {\bibinfo {volume} {127}},\
  \bibinfo {pages} {177001} (\bibinfo {year} {2021})}\BibitemShut {NoStop}%
\bibitem [{\citenamefont {Nandkishore}\ \emph {et~al.}(2012)\citenamefont
  {Nandkishore}, \citenamefont {Levitov},\ and\ \citenamefont
  {Chubukov}}]{nandkishore12np}%
  \BibitemOpen
  \bibfield  {author} {\bibinfo {author} {\bibfnamefont {R.}~\bibnamefont
  {Nandkishore}}, \bibinfo {author} {\bibfnamefont {L.~S.}\ \bibnamefont
  {Levitov}},\ and\ \bibinfo {author} {\bibfnamefont {A.~V.}\ \bibnamefont
  {Chubukov}},\ }\bibfield  {title} {\bibinfo {title} {Chiral superconductivity
  from repulsive interactions in doped graphene},\ }\href
  {https://doi.org/10.1038/nphys2208} {\bibfield  {journal} {\bibinfo
  {journal} {Nat. Phys.}\ }\textbf {\bibinfo {volume} {8}},\ \bibinfo {pages}
  {158} (\bibinfo {year} {2012})}\BibitemShut {NoStop}%
\bibitem [{\citenamefont {Kiesel}\ \emph {et~al.}(2012)\citenamefont {Kiesel},
  \citenamefont {Platt}, \citenamefont {Hanke}, \citenamefont {Abanin},\ and\
  \citenamefont {Thomale}}]{kiesel12prb}%
  \BibitemOpen
  \bibfield  {author} {\bibinfo {author} {\bibfnamefont {M.~L.}\ \bibnamefont
  {Kiesel}}, \bibinfo {author} {\bibfnamefont {C.}~\bibnamefont {Platt}},
  \bibinfo {author} {\bibfnamefont {W.}~\bibnamefont {Hanke}}, \bibinfo
  {author} {\bibfnamefont {D.~A.}\ \bibnamefont {Abanin}},\ and\ \bibinfo
  {author} {\bibfnamefont {R.}~\bibnamefont {Thomale}},\ }\bibfield  {title}
  {\bibinfo {title} {Competing many-body instabilities and unconventional
  superconductivity in graphene},\ }\href
  {https://doi.org/10.1103/PhysRevB.86.020507} {\bibfield  {journal} {\bibinfo
  {journal} {Phys. Rev. B}\ }\textbf {\bibinfo {volume} {86}},\ \bibinfo
  {pages} {020507} (\bibinfo {year} {2012})}\BibitemShut {NoStop}%
\bibitem [{\citenamefont {Yu}\ and\ \citenamefont {Li}(2012)}]{yu12prb}%
  \BibitemOpen
  \bibfield  {author} {\bibinfo {author} {\bibfnamefont {S.-L.}\ \bibnamefont
  {Yu}}\ and\ \bibinfo {author} {\bibfnamefont {J.-X.}\ \bibnamefont {Li}},\
  }\bibfield  {title} {\bibinfo {title} {Chiral superconducting phase and
  chiral spin-density-wave phase in a hubbard model on the kagome lattice},\
  }\href {https://doi.org/10.1103/PhysRevB.85.144402} {\bibfield  {journal}
  {\bibinfo  {journal} {Phys. Rev. B}\ }\textbf {\bibinfo {volume} {85}},\
  \bibinfo {pages} {144402} (\bibinfo {year} {2012})}\BibitemShut {NoStop}%
\bibitem [{\citenamefont {Kiesel}\ \emph {et~al.}(2013)\citenamefont {Kiesel},
  \citenamefont {Platt},\ and\ \citenamefont {Thomale}}]{kiesel13prl}%
  \BibitemOpen
  \bibfield  {author} {\bibinfo {author} {\bibfnamefont {M.~L.}\ \bibnamefont
  {Kiesel}}, \bibinfo {author} {\bibfnamefont {C.}~\bibnamefont {Platt}},\ and\
  \bibinfo {author} {\bibfnamefont {R.}~\bibnamefont {Thomale}},\ }\bibfield
  {title} {\bibinfo {title} {Unconventional fermi surface instabilities in the
  kagome hubbard model},\ }\href
  {https://doi.org/10.1103/PhysRevLett.110.126405} {\bibfield  {journal}
  {\bibinfo  {journal} {Phys. Rev. Lett.}\ }\textbf {\bibinfo {volume} {110}},\
  \bibinfo {pages} {126405} (\bibinfo {year} {2013})}\BibitemShut {NoStop}%
\bibitem [{\citenamefont {Wang}\ \emph {et~al.}(2013)\citenamefont {Wang},
  \citenamefont {Li}, \citenamefont {Xiang},\ and\ \citenamefont
  {Wang}}]{wang13prb}%
  \BibitemOpen
  \bibfield  {author} {\bibinfo {author} {\bibfnamefont {W.-S.}\ \bibnamefont
  {Wang}}, \bibinfo {author} {\bibfnamefont {Z.-Z.}\ \bibnamefont {Li}},
  \bibinfo {author} {\bibfnamefont {Y.-Y.}\ \bibnamefont {Xiang}},\ and\
  \bibinfo {author} {\bibfnamefont {Q.-H.}\ \bibnamefont {Wang}},\ }\bibfield
  {title} {\bibinfo {title} {Competing electronic orders on kagome lattices at
  van hove filling},\ }\href {https://doi.org/10.1103/PhysRevB.87.115135}
  {\bibfield  {journal} {\bibinfo  {journal} {Phys. Rev. B}\ }\textbf {\bibinfo
  {volume} {87}},\ \bibinfo {pages} {115135} (\bibinfo {year}
  {2013})}\BibitemShut {NoStop}%
\bibitem [{\citenamefont {Nandkishore}\ \emph {et~al.}(2014)\citenamefont
  {Nandkishore}, \citenamefont {Thomale},\ and\ \citenamefont
  {Chubukov}}]{nandkishore14prb}%
  \BibitemOpen
  \bibfield  {author} {\bibinfo {author} {\bibfnamefont {R.}~\bibnamefont
  {Nandkishore}}, \bibinfo {author} {\bibfnamefont {R.}~\bibnamefont
  {Thomale}},\ and\ \bibinfo {author} {\bibfnamefont {A.~V.}\ \bibnamefont
  {Chubukov}},\ }\bibfield  {title} {\bibinfo {title} {Superconductivity from
  weak repulsion in hexagonal lattice systems},\ }\href
  {https://doi.org/10.1103/PhysRevB.89.144501} {\bibfield  {journal} {\bibinfo
  {journal} {Phys. Rev. B}\ }\textbf {\bibinfo {volume} {89}},\ \bibinfo
  {pages} {144501} (\bibinfo {year} {2014})}\BibitemShut {NoStop}%
\bibitem [{\citenamefont {Classen}\ \emph {et~al.}(2020)\citenamefont
  {Classen}, \citenamefont {Chubukov}, \citenamefont {Honerkamp},\ and\
  \citenamefont {Scherer}}]{classen20prb}%
  \BibitemOpen
  \bibfield  {author} {\bibinfo {author} {\bibfnamefont {L.}~\bibnamefont
  {Classen}}, \bibinfo {author} {\bibfnamefont {A.~V.}\ \bibnamefont
  {Chubukov}}, \bibinfo {author} {\bibfnamefont {C.}~\bibnamefont
  {Honerkamp}},\ and\ \bibinfo {author} {\bibfnamefont {M.~M.}\ \bibnamefont
  {Scherer}},\ }\bibfield  {title} {\bibinfo {title} {Competing orders at
  higher-order van hove points},\ }\href
  {https://doi.org/10.1103/PhysRevB.102.125141} {\bibfield  {journal} {\bibinfo
   {journal} {Phys. Rev. B}\ }\textbf {\bibinfo {volume} {102}},\ \bibinfo
  {pages} {125141} (\bibinfo {year} {2020})}\BibitemShut {NoStop}%
\bibitem [{\citenamefont {Lin}\ and\ \citenamefont
  {Nandkishore}(2020)}]{lin20prb}%
  \BibitemOpen
  \bibfield  {author} {\bibinfo {author} {\bibfnamefont {Y.-P.}\ \bibnamefont
  {Lin}}\ and\ \bibinfo {author} {\bibfnamefont {R.~M.}\ \bibnamefont
  {Nandkishore}},\ }\bibfield  {title} {\bibinfo {title} {Parquet
  renormalization group analysis of weak-coupling instabilities with multiple
  high-order van hove points inside the brillouin zone},\ }\href
  {https://doi.org/10.1103/PhysRevB.102.245122} {\bibfield  {journal} {\bibinfo
   {journal} {Phys. Rev. B}\ }\textbf {\bibinfo {volume} {102}},\ \bibinfo
  {pages} {245122} (\bibinfo {year} {2020})}\BibitemShut {NoStop}%
\bibitem [{\citenamefont {Coleman}(2015)}]{colemanmbp}%
  \BibitemOpen
  \bibfield  {author} {\bibinfo {author} {\bibfnamefont {P.}~\bibnamefont
  {Coleman}},\ }\href {https://doi.org/10.1017/CBO9781139020916} {\emph
  {\bibinfo {title} {Introduction to Many-Body Physics}}}\ (\bibinfo
  {publisher} {Cambridge University Press, Cambridge, England},\ \bibinfo
  {year} {2015})\BibitemShut {NoStop}%
\bibitem [{\citenamefont {Lin}\ and\ \citenamefont
  {Nandkishore}(2018)}]{lin18prb}%
  \BibitemOpen
  \bibfield  {author} {\bibinfo {author} {\bibfnamefont {Y.-P.}\ \bibnamefont
  {Lin}}\ and\ \bibinfo {author} {\bibfnamefont {R.~M.}\ \bibnamefont
  {Nandkishore}},\ }\bibfield  {title} {\bibinfo {title} {Kohn-luttinger
  superconductivity on two orbital honeycomb lattice},\ }\href
  {https://doi.org/10.1103/PhysRevB.98.214521} {\bibfield  {journal} {\bibinfo
  {journal} {Phys. Rev. B}\ }\textbf {\bibinfo {volume} {98}},\ \bibinfo
  {pages} {214521} (\bibinfo {year} {2018})}\BibitemShut {NoStop}%
\bibitem [{\citenamefont {Lin}(2020)}]{lin20prr}%
  \BibitemOpen
  \bibfield  {author} {\bibinfo {author} {\bibfnamefont {Y.-P.}\ \bibnamefont
  {Lin}},\ }\bibfield  {title} {\bibinfo {title} {Chiral flat band
  superconductivity from symmetry-protected three-band crossings},\ }\href
  {https://doi.org/10.1103/PhysRevResearch.2.043209} {\bibfield  {journal}
  {\bibinfo  {journal} {Phys. Rev. Research}\ }\textbf {\bibinfo {volume}
  {2}},\ \bibinfo {pages} {043209} (\bibinfo {year} {2020})}\BibitemShut
  {NoStop}%
\bibitem [{\citenamefont {Fernandes}\ and\ \citenamefont
  {Millis}(2013)}]{fernandes13prl}%
  \BibitemOpen
  \bibfield  {author} {\bibinfo {author} {\bibfnamefont {R.~M.}\ \bibnamefont
  {Fernandes}}\ and\ \bibinfo {author} {\bibfnamefont {A.~J.}\ \bibnamefont
  {Millis}},\ }\bibfield  {title} {\bibinfo {title} {Nematicity as a probe of
  superconducting pairing in iron-based superconductors},\ }\href
  {https://doi.org/10.1103/PhysRevLett.111.127001} {\bibfield  {journal}
  {\bibinfo  {journal} {Phys. Rev. Lett.}\ }\textbf {\bibinfo {volume} {111}},\
  \bibinfo {pages} {127001} (\bibinfo {year} {2013})}\BibitemShut {NoStop}%
\bibitem [{\citenamefont {Fernandes}\ \emph {et~al.}(2014)\citenamefont
  {Fernandes}, \citenamefont {Chubukov},\ and\ \citenamefont
  {Schmalian}}]{fernandes14np}%
  \BibitemOpen
  \bibfield  {author} {\bibinfo {author} {\bibfnamefont {R.~M.}\ \bibnamefont
  {Fernandes}}, \bibinfo {author} {\bibfnamefont {A.~V.}\ \bibnamefont
  {Chubukov}},\ and\ \bibinfo {author} {\bibfnamefont {J.}~\bibnamefont
  {Schmalian}},\ }\bibfield  {title} {\bibinfo {title} {What drives nematic
  order in iron-based superconductors?},\ }\href
  {https://doi.org/10.1038/nphys2877} {\bibfield  {journal} {\bibinfo
  {journal} {Nat. Phys.}\ }\textbf {\bibinfo {volume} {10}},\ \bibinfo {pages}
  {97} (\bibinfo {year} {2014})}\BibitemShut {NoStop}%
\bibitem [{\citenamefont {Chen}\ \emph {et~al.}(2020)\citenamefont {Chen},
  \citenamefont {Maiti}, \citenamefont {Fernandes},\ and\ \citenamefont
  {Hirschfeld}}]{chen20prb}%
  \BibitemOpen
  \bibfield  {author} {\bibinfo {author} {\bibfnamefont {X.}~\bibnamefont
  {Chen}}, \bibinfo {author} {\bibfnamefont {S.}~\bibnamefont {Maiti}},
  \bibinfo {author} {\bibfnamefont {R.~M.}\ \bibnamefont {Fernandes}},\ and\
  \bibinfo {author} {\bibfnamefont {P.~J.}\ \bibnamefont {Hirschfeld}},\
  }\bibfield  {title} {\bibinfo {title} {Nematicity and superconductivity:
  Competition versus cooperation},\ }\href
  {https://doi.org/10.1103/PhysRevB.102.184512} {\bibfield  {journal} {\bibinfo
   {journal} {Phys. Rev. B}\ }\textbf {\bibinfo {volume} {102}},\ \bibinfo
  {pages} {184512} (\bibinfo {year} {2020})}\BibitemShut {NoStop}%
\bibitem [{\citenamefont {Abanov}\ \emph {et~al.}(2003)\citenamefont {Abanov},
  \citenamefont {Chubukov},\ and\ \citenamefont {Schmalian}}]{abanov03advp}%
  \BibitemOpen
  \bibfield  {author} {\bibinfo {author} {\bibfnamefont {A.}~\bibnamefont
  {Abanov}}, \bibinfo {author} {\bibfnamefont {A.~V.}\ \bibnamefont
  {Chubukov}},\ and\ \bibinfo {author} {\bibfnamefont {J.}~\bibnamefont
  {Schmalian}},\ }\bibfield  {title} {\bibinfo {title} {Quantum-critical theory
  of the spin-fermion model and its application to cuprates: Normal state
  analysis},\ }\href {https://doi.org/10.1080/0001873021000057123} {\bibfield
  {journal} {\bibinfo  {journal} {Adv. Phys.}\ }\textbf {\bibinfo {volume}
  {52}},\ \bibinfo {pages} {119} (\bibinfo {year} {2003})}\BibitemShut
  {NoStop}%
\bibitem [{\citenamefont {Peotta}\ and\ \citenamefont
  {T\"orm\"a}(2015)}]{peotta15nc}%
  \BibitemOpen
  \bibfield  {author} {\bibinfo {author} {\bibfnamefont {S.}~\bibnamefont
  {Peotta}}\ and\ \bibinfo {author} {\bibfnamefont {P.}~\bibnamefont
  {T\"orm\"a}},\ }\bibfield  {title} {\bibinfo {title} {Superfluidity in
  topologically nontrivial flat bands},\ }\href
  {https://doi.org/10.1038/ncomms9944} {\bibfield  {journal} {\bibinfo
  {journal} {Nat. Commun.}\ }\textbf {\bibinfo {volume} {6}},\ \bibinfo {pages}
  {8944} (\bibinfo {year} {2015})}\BibitemShut {NoStop}%
\bibitem [{\citenamefont {Liang}\ \emph {et~al.}(2017)\citenamefont {Liang},
  \citenamefont {Vanhala}, \citenamefont {Peotta}, \citenamefont {Siro},
  \citenamefont {Harju},\ and\ \citenamefont {T\"orm\"a}}]{liang17prb}%
  \BibitemOpen
  \bibfield  {author} {\bibinfo {author} {\bibfnamefont {L.}~\bibnamefont
  {Liang}}, \bibinfo {author} {\bibfnamefont {T.~I.}\ \bibnamefont {Vanhala}},
  \bibinfo {author} {\bibfnamefont {S.}~\bibnamefont {Peotta}}, \bibinfo
  {author} {\bibfnamefont {T.}~\bibnamefont {Siro}}, \bibinfo {author}
  {\bibfnamefont {A.}~\bibnamefont {Harju}},\ and\ \bibinfo {author}
  {\bibfnamefont {P.}~\bibnamefont {T\"orm\"a}},\ }\bibfield  {title} {\bibinfo
  {title} {Band geometry, berry curvature, and superfluid weight},\ }\href
  {https://doi.org/10.1103/PhysRevB.95.024515} {\bibfield  {journal} {\bibinfo
  {journal} {Phys. Rev. B}\ }\textbf {\bibinfo {volume} {95}},\ \bibinfo
  {pages} {024515} (\bibinfo {year} {2017})}\BibitemShut {NoStop}%
\bibitem [{\citenamefont {Hu}\ \emph {et~al.}(2019)\citenamefont {Hu},
  \citenamefont {Hyart}, \citenamefont {Pikulin},\ and\ \citenamefont
  {Rossi}}]{hu19prl}%
  \BibitemOpen
  \bibfield  {author} {\bibinfo {author} {\bibfnamefont {X.}~\bibnamefont
  {Hu}}, \bibinfo {author} {\bibfnamefont {T.}~\bibnamefont {Hyart}}, \bibinfo
  {author} {\bibfnamefont {D.~I.}\ \bibnamefont {Pikulin}},\ and\ \bibinfo
  {author} {\bibfnamefont {E.}~\bibnamefont {Rossi}},\ }\bibfield  {title}
  {\bibinfo {title} {Geometric and conventional contribution to the superfluid
  weight in twisted bilayer graphene},\ }\href
  {https://doi.org/10.1103/PhysRevLett.123.237002} {\bibfield  {journal}
  {\bibinfo  {journal} {Phys. Rev. Lett.}\ }\textbf {\bibinfo {volume} {123}},\
  \bibinfo {pages} {237002} (\bibinfo {year} {2019})}\BibitemShut {NoStop}%
\bibitem [{\citenamefont {Xie}\ \emph {et~al.}(2020)\citenamefont {Xie},
  \citenamefont {Song}, \citenamefont {Lian},\ and\ \citenamefont
  {Bernevig}}]{xie20prl}%
  \BibitemOpen
  \bibfield  {author} {\bibinfo {author} {\bibfnamefont {F.}~\bibnamefont
  {Xie}}, \bibinfo {author} {\bibfnamefont {Z.}~\bibnamefont {Song}}, \bibinfo
  {author} {\bibfnamefont {B.}~\bibnamefont {Lian}},\ and\ \bibinfo {author}
  {\bibfnamefont {B.~A.}\ \bibnamefont {Bernevig}},\ }\bibfield  {title}
  {\bibinfo {title} {Topology-bounded superfluid weight in twisted bilayer
  graphene},\ }\href {https://doi.org/10.1103/PhysRevLett.124.167002}
  {\bibfield  {journal} {\bibinfo  {journal} {Phys. Rev. Lett.}\ }\textbf
  {\bibinfo {volume} {124}},\ \bibinfo {pages} {167002} (\bibinfo {year}
  {2020})}\BibitemShut {NoStop}%
\bibitem [{\citenamefont {Lin}\ and\ \citenamefont {Hsiao}(2021)}]{lin21prbdh}%
  \BibitemOpen
  \bibfield  {author} {\bibinfo {author} {\bibfnamefont {Y.-P.}\ \bibnamefont
  {Lin}}\ and\ \bibinfo {author} {\bibfnamefont {W.-H.}\ \bibnamefont
  {Hsiao}},\ }\bibfield  {title} {\bibinfo {title} {Dual haldane sphere and
  quantized band geometry in chiral multifold fermions},\ }\href
  {https://doi.org/10.1103/PhysRevB.103.L081103} {\bibfield  {journal}
  {\bibinfo  {journal} {Phys. Rev. B}\ }\textbf {\bibinfo {volume} {103}},\
  \bibinfo {pages} {L081103} (\bibinfo {year} {2021})}\BibitemShut {NoStop}%
\bibitem [{\citenamefont {Li}\ and\ \citenamefont {Haldane}(2018)}]{li18prl}%
  \BibitemOpen
  \bibfield  {author} {\bibinfo {author} {\bibfnamefont {Y.}~\bibnamefont
  {Li}}\ and\ \bibinfo {author} {\bibfnamefont {F.~D.~M.}\ \bibnamefont
  {Haldane}},\ }\bibfield  {title} {\bibinfo {title} {Topological nodal cooper
  pairing in doped weyl metals},\ }\href
  {https://doi.org/10.1103/PhysRevLett.120.067003} {\bibfield  {journal}
  {\bibinfo  {journal} {Phys. Rev. Lett.}\ }\textbf {\bibinfo {volume} {120}},\
  \bibinfo {pages} {067003} (\bibinfo {year} {2018})}\BibitemShut {NoStop}%
\bibitem [{\citenamefont {Berg}\ \emph {et~al.}(2012)\citenamefont {Berg},
  \citenamefont {Metlitski},\ and\ \citenamefont {Sachdev}}]{berg12s}%
  \BibitemOpen
  \bibfield  {author} {\bibinfo {author} {\bibfnamefont {E.}~\bibnamefont
  {Berg}}, \bibinfo {author} {\bibfnamefont {M.~A.}\ \bibnamefont
  {Metlitski}},\ and\ \bibinfo {author} {\bibfnamefont {S.}~\bibnamefont
  {Sachdev}},\ }\bibfield  {title} {\bibinfo {title}
  {Sign-problem\&\#x2013;free quantum monte carlo of the onset of
  antiferromagnetism in metals},\ }\href
  {https://doi.org/10.1126/science.1227769} {\bibfield  {journal} {\bibinfo
  {journal} {Science}\ }\textbf {\bibinfo {volume} {338}},\ \bibinfo {pages}
  {1606} (\bibinfo {year} {2012})}\BibitemShut {NoStop}%
\bibitem [{\citenamefont {Rodriguez}(2021)}]{rodriguez21prb}%
  \BibitemOpen
  \bibfield  {author} {\bibinfo {author} {\bibfnamefont {J.~P.}\ \bibnamefont
  {Rodriguez}},\ }\bibfield  {title} {\bibinfo {title} {Superconductivity by
  hidden spin fluctuations in electron-doped iron selenide},\ }\href
  {https://doi.org/10.1103/PhysRevB.103.184513} {\bibfield  {journal} {\bibinfo
   {journal} {Phys. Rev. B}\ }\textbf {\bibinfo {volume} {103}},\ \bibinfo
  {pages} {184513} (\bibinfo {year} {2021})}\BibitemShut {NoStop}%
\bibitem [{\citenamefont {Tazai}\ \emph {et~al.}(2022)\citenamefont {Tazai},
  \citenamefont {Yamakawa}, \citenamefont {Onari},\ and\ \citenamefont
  {Kontani}}]{tazai2sa}%
  \BibitemOpen
  \bibfield  {author} {\bibinfo {author} {\bibfnamefont {R.}~\bibnamefont
  {Tazai}}, \bibinfo {author} {\bibfnamefont {Y.}~\bibnamefont {Yamakawa}},
  \bibinfo {author} {\bibfnamefont {S.}~\bibnamefont {Onari}},\ and\ \bibinfo
  {author} {\bibfnamefont {H.}~\bibnamefont {Kontani}},\ }\bibfield  {title}
  {\bibinfo {title} {{Mechanism of exotic density-wave and beyond-Migdal
  unconventional superconductivity in kagome metal AV$_3$Sb$_5$ (A = K, Rb,
  Cs)}},\ }\href {https://doi.org/10.1126/sciadv.abl4108} {\bibfield  {journal}
  {\bibinfo  {journal} {Sci. Adv.}\ }\textbf {\bibinfo {volume} {8}},\ \bibinfo
  {pages} {eabl4108} (\bibinfo {year} {2022})}\BibitemShut {NoStop}%
\end{thebibliography}%

\end{document}